\documentclass[a4paper,11pt]{article}
\pdfoutput=1
\usepackage{jcappub} 
\usepackage[utf8]{inputenc} 
\usepackage[T1]{fontenc}
\usepackage[titletoc,title]{appendix}
\newcommand{\beq}{\begin{equation}}
\newcommand{\eeq}{\end{equation}}
\usepackage{graphicx}
\graphicspath{{./Figures/}}
\usepackage{amsmath}
\usepackage{relsize}
\usepackage{xcolor}
\usepackage{soul}
\usepackage{MnSymbol}
\usepackage{mathrsfs}
\usepackage{subcaption}
\usepackage{mwe}
\usepackage{gensymb}
\usepackage{slashed}

\begin{document}

\title{\boldmath Constraints on Dark Matter with a moderately large and velocity-dependent DM-nucleon cross-section}
\author{M. Shafi Mahdawi}
\author{and Glennys R. Farrar}
\affiliation{Center for Cosmology and Particle Physics, Department of Physics, New York University, \\4 Washington Place, New York, NY 10003, USA}
\emailAdd{shafi.mahdawi@nyu.edu}
\emailAdd{gf25@nyu.edu}

\abstract{We derive constraints on a possible velocity-dependent DM-nucleon scattering cross section, for Dark Matter in the 10 MeV -- 100 GeV mass range, using the XQC, DAMIC, and CRESST 2017 Surface Run experiments. We report the limits on cross sections of the form $\sigma=\sigma_0\,v^n$, for a range of velocity dependencies with $n\in\{-4,-2,-1,0,1,2\}$. We point out the need to measure the  efficiency with which nuclear recoil energy in the sub-keV range thermalizes, rather than being stored as Frenkel pairs in the semi-conductor lattice.  The possibility of a significant inefficiency leaves open a considerable ``hole'' in the limits for mass in the $\sim$ 0.2 -- 2 GeV range, which XQC and CRESST can potentially fill when the thermalization efficiency is measured. We call attention to the asymmetry between a conventional lower limit cross section and the ``upper-reach cross section'' imposed by attenuation in an overburden -- an upper boundary being extremely sharp but quite insensitive to the statistics of the experiment.  Considering the recent interest to use dark matter-baryon interaction with velocity dependence $n=-4$ to explain the EDGES 21 cm anomaly, we also derive the limits on milli-charged DM that scatters off protons and electrons under a Coulomb-like interaction.  We find that much but not all of the region of interest for the EDGES anomaly can be excluded.  }
\maketitle
\flushbottom
\section{Introduction}
Despite well established astrophysical and cosmological evidence for Dark Matter (DM), its particle nature remains a mystery~\cite{Patrignani:2016xqp}. Weakly-interacting massive particle (WIMP), the most hunted DM candidate, has evaded all DM searches. Hadronically-interacting DM\footnote{ This refers to interaction between DM and nucleon, with cross section in the range seen in hadron interactions, i.e., $10^{-30} - 10^{-23} \, {\rm cm}^2$; the term strongly-interacting DM can be confused with SIDM which refers to self-interacting DM \cite{Spergel:1999mh}.} is an alternative DM candidate that deserves studying in greater detail. 

Stringent constraints on hadronically-interacting DM exist from ground-based direct detection experiments~\cite{Emken, nuDavis, bounds, nuKavanagh, Emken2018}, balloon and satellite experiments~\cite{Wandelt:2000ad, RICH1987}, the XQC rocket experiment~\cite{McCammon2002,bounds}, CMB and Lyman-$\alpha$ power spectrum~\cite{Dvorkin:2013cea,Gluscevic2017, Xu2018}, molecular spectroscopy~\cite{Fichet:2017bng}, and LHC searches~\cite{Daci:2015hca}.  Recently, an analysis of the DM atmosphere surrounding Earth has provided powerful new constraints for a limited DM mass range, $\approx 0.6-6$ GeV, from the orbital decay of the HST and longevity of the LHC beam, storage times of liquid cryogens, and the Earth's thermal conductivity~\cite{nfm18}.

In this work, we derive the constraints on hadronically-interacting Dark Matter using the observed event energies of the direct detection experiments, considering DM-nucleon cross section with power-law velocity-dependence, i.e. $\sigma=\sigma_0\,v^n$ where $v$ is the velocity of DM in the lab frame and $c=$1. Considering the range of velocity dependencies $n\in\{-4,-2,-1,0,1,2\}$ allows us to probe the underlying particle physics of DM. These velocity dependencies are well-motivated, e.g., $n=-4$ occurs in milli-charged DM~\cite{Holdom:1985ag,Chun:2010ve}, $n=\pm2$ appears in a DM with electric and/or magnetic dipole moment~\cite{Sigurdson:2004zp}, and other power-law velocity-dependencies can occur in a Yukawa potential~\cite{Xingchen2018, Tulin2013, ArkaniHamed:2008qn, Buckley:2009in}.  We make the standard assumption that cross sections on different nuclei are related as in Born approximation.

We also point out that for the nuclear recoil energies encountered in constraining DM masses $\lesssim$ 2 GeV, a significant fraction of nuclear recoil energy $E_{\rm{nr}}$ can be stored in interstitial lattice defects.  This reduces the efficiency, $\epsilon_{\rm{th}}$, with which $E_{\rm{nr}}$ is converted to the thermal energy measured by XQC, DAMIC and CRESST. As discussed in appendix \ref{ap:effth}, $\epsilon_{\rm{th}}$ may be only a few percent. We therefore analyze XQC with $\epsilon_{\rm{th}}=$ 0.02, 0.1 and 1, to show the sensitivity to this parameter and illustrate the importance of measuring it.  

Recently, the EDGES collaboration~\cite{Bowman2018} reported a stronger than expected absorption signal corresponding to the 21-cm-wavelength transition of atomic hydrogen around redshift $z=17$. To explain a colder than expected baryonic matter during cosmic dawn, based on earlier studies~\cite{Tashiro:2014tsa, Munoz:2015bca}, ref.~\cite{Barkana2018} proposed a DM-nucleon interaction with velocity dependence of the form $\sigma\propto v^{-4}$. This possibility has been further studied by considering the constraints on milli-charged DM~\cite{Munoz:2018pzp, Berlin:2018sjs, Barkana:2018qrx, Fraser:2018acy, Munoz:2018jwq}, hadronically interacting DM~\cite{nfm18}, imprints of such a DM model on the baryonic acoustic oscillation (BAO)~\cite{Fialkov:2018xre}, and several other cosmological constraints in~\cite{Slatyer:2018aqg}. In this work, beside reporting the bounds for velocity-dependence $n=-4$ in the heavy-mediator limit, we report the bounds on milli-charged DM~\cite{Holdom:1985ag, Chun:2010ve} with transfer cross section having $n=-4$ velocity-dependence. Our result further constrains milli-charged DM~\cite{Prinz:1998ua,Davidson:2000hf, Dubovsky:2003yn, Chuzhoy:2008zy, Dolgov:2013una, McDermott:2010pa, Vogel:2013raa, Haas:2014dda, Essig:2017kqs, Chang:2018rso, Crisler:2018gci}.

This work is organized as follows. We introduce the direct detection experiments that we use in this work in section~\ref{Data_Analysis}. Milli-charged DM is reviewed briefly in section~\ref{mCP_theory}. The steps of our Monte-Carlo simulation to calculate the nuclear energy-loss of a DM particle with a velocity-dependent interaction in an overburden are discussed in section~\ref{MC_description}. Details of the Monte-Carlo simulation to calculate the total energy deposited in the XQC calorimeter in the multiple-scatterings case are discussed in section~\ref{multiple_scatterings}. Our results are presented in section~\ref{result_vb}. Complementary details are presented in appendices~\ref{ap:effth} -- \ref{XQC_detail}.
\section{Data and analysis}\label{Data_Analysis}
We used the observed energy-deposit spectra from the XQC~\cite{McCammon2002}, DAMIC~\cite{Barreto2012264}, and CRESST 2017 surface run~\cite{nu1} experiments to find the bounds on milli-charged DM and strongly interacting DM, for a power-law velocity dependent DM-nucleon cross section. 
\subsection{CRESST 2017 surface run}	
The CRESST 2017 surface run\footnote{ a prototype detector which was developed for the $\nu$-cleus experiment.} (CSR) apparatus~\cite{nu1,nu2,nu3}, the cryogenic detector operated by the CRESST collaboration above ground at Max-Planck-Institute for Physics in Munich, reached down to a nuclear recoil energy threshold $E_{\rm{nr}}^{th}=19.7$ eV. The CSR detector, made of $\rm{Al_2O_3}$, accumulated a total exposure of 0.046 g$\cdot$days during its net live-time of 2.27 h, and observed a total of 511 events in the nuclear recoil energy range from 19.7 eV to 600 eV~\cite{nu1}. The CSR experiment with minimal shielding of $\sim$30 cm concrete is an ideal experiment for constraining hadronically-interacting DM. 
\par
Starkman, Gould, Esmailzadeh and Dimopolous~\cite{Starkman} (SGED) proposed a continuous energy-loss and vertical propagation approximation, to estimate the maximum cross section for which a given experiment is sensitive to strongly interacting DM particles. Reference~\cite{method} discusses in detail the limitations of this approximation. Using the CSR observed spectrum, \cite{nuDavis,nuKavanagh} used the SGED method to find the bounds on hadronically-interacting DM with velocity-independent DM-nucleon cross section and \cite{Emken2018} used Monte-Carlo simulation to find the limits on a velocity-independent DM-nucleon cross section using CSR. 
\par
We use an importance sampling Monte-Carlo simulation~\cite{bounds, method}, discussed in detail in section~\ref{MC_description}, which takes into account the effect of Earth's reflection of hadronically-interacting DM particles (see appendix~\ref{DMATIS_mod} for a detailed discussion of this effect), to find the limits in the 150 MeV -- 100 GeV mass range where we expect to see the largest deviation from the SGED approximation~\cite{method}. To obtain more conservative bounds\footnote{ No data quality cuts are applied to the observed spectrum of the CSR experiment.}, we use the total 511 number of events observed by the CSR experiment in the 19.7 eV -- 600 eV nuclear recoil energy range, to find the upper-cross-section reach of this experiment. The 90\% limit on the allowed (by CSR) cross section is defined to be the value of $\sigma_0$ for which the expected total number of events is 547. (The equivalent 90\% CL upper limit on 511 observed total number of events being 547.) 

We generalized the SGED approximation~\cite{Starkman, method}, suitable in the limit of many interactions with small-deflection angles and essentially continuous energy-loss, to analytically calculate energy-loss of (a) DM particles with power-law velocity-dependent DM-nucleon cross section (see appendix~\ref{SGEDvel}) and (b) milli-charged DM (see appendix~\ref{SGEDcol}). We present the limits calculated using this analytic method and compare them with the limits calculated using the Monte-Carlo simulation.

\subsection{DAMIC }
The DAMIC --- Dark Matter In CCDs~\cite{Barreto2012264}  --- experiment was operated at a depth of 106.7 meters underground in the NuMI near-detector hall at Fermilab. The DAMIC detector, made of silicon and shielded by 6-inch lead, accumulated a total exposure of 107 g$\cdot$days from June 2010 to May 2011, and observed a total of 106 events with an ionization signal between 40 $\rm{eV_{ee}}$ (eV electron equivalent energy) and 2 k$\rm{eV_{ee}}$.
\par
Reference~\cite{bounds} used an importance sampling Monte-Carlo simulation to find the constraints on velocity-independent DM-nucleon cross section. Here, we extend that work to cover a range of power-law velocity-dependent cross sections by using the importance sampling Monte-Carlo simulation to calculate the bounds on hadronically-interacting DM. 
\par
The DAMIC collaboration used three selection cuts to separate DM-induced nuclear events from background events. To measure the efficiency of these selection cuts, the detector was exposed to a $^{252}$Cf neutron emitting source. By taking the ratio of the selected events to the expected events without selection cuts, an efficiency factor for selecting nuclear recoils is determined (see FIG.9 in~\cite{Barreto2012264}). In calculating the expected number of events for DAMIC, we weight each event by this efficiency factor. For this, we use the recent measurements of quenching factor in silicon in~\cite{Chavarria:2016xsi} to calculate the equivalent ionization energy for the nuclear recoil energy of each event. The 90\% limit on the allowed (by DAMIC) cross section is defined to be the value of $\sigma_0$ for which the expected total number of events is 123. (The equivalent 90\% CL upper limit on 106 observed total number of events being 123.)
\subsection{XQC}
XQC --- the X-ray Quantum Calorimeter~\cite{McCammon2002} --- was an X-ray detector aboard a sounding-rocket launched on March 28, 1999. Each of the 34 XQC calorimeters was composed of a 0.96 $\mu$m film of HgTe mounted on a 14 $\mu$m substrate of Si. During its net live-time of 100.7 seconds at height of 201 km above the ground, XQC accumulated a total exposure of $\sim\rm{1.8\times10^{-6}}$ g$\cdot$days and observed a total of 587 events above its threshold nuclear recoil energy of 29 eV \footnote{ See appendix~\ref{XQC_detail} for details of the XQC detector mass composition, binning of its spectrum, and  its exposure time which depends on the measured thermal energy.}.
\par
Taking the observed spectrum of the XQC experiment at face value,~\cite{bounds} showed that the rocket body did not shield the detector from the DM flux in 300 MeV -- 100 GeV mass range and obtained an-order-of-magnitude stronger limits. In this work, we extend the lower reach of XQC to masses as low as 10 MeV by considering the possibility of multiple interactions in the XQC detector such that the total energy deposit exceeds the threshold. In section~\ref{multiple_scatterings}, we discuss the details of our Monte-Carlo simulation that we use to calculate the expected spectrum for XQC in the multiple scatterings case.
\par
XQC uses a quantum micro-calorimeter to measure the thermal energy deposited by X-ray hitting the detector. As discussed in Appendix \ref{ap:effth},
the XQC detector was calibrated using an X-ray emitting source and may not be fully sensitive in measuring DM-induced nuclear recoil energies due to the nuclear recoil energy becoming stored in lattice defects (Frenkel pairs) instead of thermalization. The efficiency could be as low as 2\% or less. Therefore, in this work, we adopt a thermalization efficiency factor $\epsilon_{\rm{th}}=$ 0.02 to find more conservative bounds on the DM-nucleon cross section. For comparison, we also report the limits for $\epsilon_{\rm{th}}=$ 0.1 and 1. 
\par
To make our calculation of the XQC limits as general as possible, for those DM particles that pass through the aluminum body of the rocket, we simulate their propagation in this aluminum layer with thickness of 3.7 cm to find their velocity before hitting the XQC detector. Then, we include these DM particles along with the ones that come through the opening angle to find the full velocity distribution of DM particles before hitting the XQC detector. By considering the possibility of DM particles having multiple scatterings in the detector, we calculate the expected nuclear recoil energy spectrum in XQC. Then, we calculate the thermal response by using $E_T=\epsilon_{\rm{th}}\,E_{\rm{nr}}$, where $E_T$ and $E_{\rm{nr}}$ are the thermal and nuclear recoil energies respectively. We bin the events into the thermal energy bins that are given in table~\ref{t_XQC_spec_kappa}. The 90\% CL DM-nucleon cross section is calculated using $\chi^2$ figure-of-merit (described in~\cite{Erickcek2007,bounds}) which exploits the shape of the observed energy-deposit spectrum\footnote{ Considering the shape of spectrum to find the upper reach limits for CSR and DAMIC does not change the limits due to the number of DM particles with sufficient energy being strongly dependent on the number of scatterings and hence on the strength of the DM-nucleon interaction.}.
\section{Milli-charged DM}\label{mCP_theory}
In the milli-charged DM model~\cite{Holdom:1985ag}, there is a dark photon field $X_\mu$ that couples to DM $\chi$, a Dirac fermion field with mass $m$ and charge $g_{D}$ under a dark photon, which mixes with the Standard Model (SM) photon $A_\mu$. The Lagrangian for this model is
\beq
\label{eLagMilli}
\mathcal{L}=\mathcal{L}_{\rm{SM}}+\overline{\chi}(i\,\gamma^\mu \partial_\mu+g_D\,\gamma^\mu X_\mu-m)\,\chi-\frac{1}{4}X^{\mu\nu}X_{\mu\nu}-\frac{\kappa}{2}X^{\mu\nu}F_{\mu\nu}\;,
\eeq
where $X_{\mu\nu}$ and $F_{\mu\nu}$ are the field strength tensors of the dark photon and the standard model (SM) photon fields. 

The kinematic mixing term can be eliminated by field redefinition $X_\mu\rightarrow X_\mu-\kappa\,A_\mu$. Consequently, DM field couples to SM photon and obtains electric milli-charge $\epsilon\,e\equiv\kappa\,g_D$
\beq
\label{eLagMilli2}
\mathcal{L}=\mathcal{L}_{SM}+\overline{\chi}(i\,\gamma^\mu \partial_\mu+g_D\,\gamma^\mu X_\mu- \epsilon\,e\,\gamma^\mu\,A_\mu-m)\,\chi-\frac{1}{4}X^{\mu\nu}X_{\mu\nu}\;.
\eeq
This introduces a Coulomb-like force between DM and electrically-charged SM particles. The differential DM-nucleus cross section of a milli-charged DM with velocity $v$ is~\cite{Foot:2011pi, Fornengo:2011sz,Lee:2015qva}:
\beq\begin{split}
	\label{edsAdErMilli}
	\frac{d\,\sigma_A}{d\,E_{\rm{nr}}}&=\,\frac{2\,\pi\,Z_A^2\,\epsilon^2\alpha^2}{m_A\,v^2\,E_{\rm{nr}}^2} \,F_A^2(E_{\rm{nr}}) \;,
\end{split}
\eeq
where $\alpha$ is the SM fine structure, $m_A$ is the mass of the nucleus and $Z_A$ is the charge number of the nucleus. An analytical expression which is proposed by Helm~\cite{Helm} for the nuclear form factor $F_A(E_{\rm{nr}})$ is
\beq
\label{e_FormFactor}
F_A(E_{\rm{nr}})=F(qr_A)=3\,\frac{\sin(qr_A)-qr_A\cos(qr_A)}{(qr_A)^3}\,e^{-(qs)^2/2}\;,
\eeq
where $r_A$ is the effective nuclear radius. $q\equiv\sqrt{2\,m_A\,E_{\rm{nr}}}$ is the momentum transfer. The effective nuclear radius $r_A$ can be approximately found by fitting the muon scattering data to a Fermi distribution~\cite{fricke1995}
\beq
\label{e_NuclearRadius}
r_A^2=c^2+\frac{7}{3}\pi^2a^2-5s^2\;,
\eeq
with parameters: $c\simeq(1.23A^{1/3}-0.6)$ fm, $a\simeq0.52$ fm, and $s=0.9$ fm.  

The IR divergence of the above expression is regularized at $E_{\rm{nr},A}^{\,\rm {screen}}\equiv(\alpha\,m_e)^2/2\,m_A$ due to screening of Coulomb-like force at this energy scale by electrons. The total DM-nucleus cross section is
\beq\begin{split}
	\label{esigmaAMilli}
	\sigma_A(v)\equiv\,\int^{E_{\rm{nr, A}}^{\rm{max}}}_{E_{\rm{nr, A}}^{\,\rm {screen}}}\frac{d\,\sigma_A}{d\,E_{\rm{nr}}}\,d\,E_{\rm{nr}}= \, \frac{4\,\pi\,Z_A^2\,\epsilon^2\,\alpha^2}{v^2}\,\left( \frac{1}{m_e^2\,\alpha^2}-\frac{1}{4\,\mu_A^2\,v^2}\right) \;,
\end{split}
\eeq
where $E_{\rm{nr, A}}^{\,\rm{max}}=(2\,\mu_A\,v)^2/2\,m_A$ is the maximum recoil energy. In derivation of eq.~\eqref{esigmaAMilli}, we neglected the nuclear form factor due to smallness of the momentum transfer in Coulomb-like scattering. Eq.~\eqref{esigmaAMilli} is only valid for $v$ large enough that the expression is non-negative.

The DM-nucleus transfer cross section, which is commonly used to parametrize the energy transport~\cite{Tulin2013}, is another quantity of interest for us as we consider a process in which DM particles lose part of their energy to nuclei in an overburden
\beq\begin{split}
	\label{esigmaTAMilli}
	\sigma_A^T(v)\equiv\,\int^{E_{\rm{nr}}^{\rm{max}}}_{E_{\rm{nr}}^{\,\rm {screen}}}(1-\cos\xi_{\rm{CM}})\frac{d\,\sigma_A}{d\,E_{\rm{nr}}}\,d\,E_{\rm{nr}}= \, \frac{4\,\pi\,Z_A^2\,\epsilon^2\,\alpha^2}{\mu_A^2\,v^4}\,\ln\left( \frac{2\,\mu_A\,v}{\alpha\,m_e}\right)  \;,
\end{split}
\eeq
where $\xi_{\rm{CM}}$ is the scattering angle in the center-of-mass frame.  (The transfer cross-section is so-named because the energy deposit in each collision $\sim (1-\cos\xi_{\rm{CM}})$.  For a forward-backward symmetric differential cross section there is no difference between the transfer cross section and the total cross section, but in more general cases the transfer cross section is often more useful than the scattering cross section.)

Correspondingly, the DM-nucleon transfer cross section for milli-charged DM is
\beq\begin{split}
	\label{esigmaTpMilli}
	\sigma_p^T(v)&= \, \frac{4\,\pi\,\epsilon^2\,\alpha^2}{\mu_p^2\,v^4}\,\ln\left( \frac{2\,\mu_p\,v}{\alpha\,m_e}\right)  \\ 
	&\equiv \sigma^*_T(v)\,v^{-4}\;.
\end{split}
\eeq 

\section{Monte-Carlo Simulation for CSR and DAMIC}\label{MC_description}
In this section we describe our method to calculate the impact of an overburden on the DM flux at the detector.  
\subsection{Preliminaries}
The expected differential number of events induced by DM-nucleus scatterings, in the limit of single scattering in the target, is
\beq\begin{split}
	\label{edNdEr_gen}
	\frac{dN}{dE_{\rm{nr}}}=&\sum_{A}\,t_e\,N_{A}\,\frac{\rho}{m}\int_{v_{\rm{min}}(E_{\rm{nr}},A)} \frac{d\,\sigma_{A}}{d\,E_{\rm{nr}}}\,v\,f(\vec{v},\vec{v}_{det})\,d^3v\;,
\end{split}\eeq
where $t_e$ is exposure time reported by the experiment\footnote{ $t_e$ contains experiments detection efficiency factors.}, $N_{A}\equiv M_{A}/m_{A}$ is the number of nuclei of mass number $A$ in the target, $\rho=$ 0.3 GeV$\cdot\,\rm{cm^{-3}}$ is the DM local mass density\footnote{ Milli-charged DM particles are argued to be evacuated from the Galactic disk by supernova explosions and magnetic field in the Milky Way~\cite{Chuzhoy:2008zy,McDermott:2010pa}. In the case of milli-charged DM, we find the limits by taking the DM mass density in the detector frame to be $f\,\cdot\rho$ where $f$ is the fraction of milli-charged DM particles that remain in the disk.}, $m$ is the mass of each DM particle, $v_{\rm{min}}(E_{\rm{nr}},A)\equiv\frac{\sqrt{2\,m_{A}\,E_{\rm{nr}}}}{2\,\mu_{A}}$ is the minimum speed that a DM particle needs in order to deposit the recoil energy $E_{\rm{nr}}$, $\frac{d\,\sigma_{A}}{d\,E_{\rm{nr}}}$ is the differential DM-nucleus cross section, and $f(\vec{v},\vec{v}_{det})$ is the unit-normalized velocity distribution of DM particles in the detector rest frame.

Interactions strong enough between DM and nuclei in an overburden\footnote{ According to ref~\cite{Kouvaris2014}, the electronic energy-loss is negligible in comparison to nuclear energy-loss for the case of milli-charged DM; therefore in the present analysis we consider only nuclear energy-loss in the overburden.} --- which could be the Earth's atmosphere, the Earth volume, or the experimental shielding --- significantly modify the velocity distribution of DM particles in the Earth rest frame~\cite{Starkman, ZF_window, Kouvaris2014, Kavanagh:2016pyr, Emken2018, bounds,method}.  We characterize the 
 \textit{speed} distribution of \textit{capable}\footnote{ We denote as ``capable'', those particles with enough energy to potentially trigger the detector.} DM particles at the detector, $f(v,\,m,\,\sigma)$, as follows
\beq\begin{split}
	\label{f_speed_unitNormalized_strong}
	f(v,\,m,\,\sigma)=\int\,v^2\,f(\vec{v},\vec{v}_{det},\,m,\,\sigma)\,d\,\Omega=\frac{1}{2}\,\eta(m)\,a_c(m,\,\sigma)\,f_c(v,\,m,\,\sigma)\;,
\end{split}\eeq
where $f_c(v,\,m,\,\sigma)$ is the unit-normalized speed
distribution of capable DM particles at the detector and $\eta(m)$ is the fraction of capable DM particles before entering the overburden. $a_c(m,\,\sigma)$ is thus the attenuation parameter (the ratio of the number of capable DM particles at the detector to the number of capable DM particles before entering the overburden) which is normalized to one before entering the overburden. The factor of one-half in eq.~\eqref{f_speed_unitNormalized_strong} is due the shielding of DM particles entering the overburden from below the horizon\footnote{ This assumes that the time-averaged initial zenith angle of DM particles is isotropic.  It is an excellent approximation in finding the cross section limits for both CSR and DAMIC, due to the extremely large attenuation factors that are being probed by each of these two experiments.  E.g., a factor-2 difference in number of events has a negligible impact on the inferred cross section limit.  We quantify the accuracy of this assumption in the next subsection.
}.
\subsection{Monte-Carlo simulation}
In this section, we review the basics of our Monte-Carlo simulation implemented to calculate attenuation parameter $a_c(m,\,\sigma)$ and the speed distribution of capable DM particles at the CSR detector,  $f_c(v,\,m,\,\sigma)$, by considering the Earth’s atmosphere and the concrete layer surrounding CSR as the main shielding layers\footnote{ Our simulation and the result presented here includes the shielding effect of the concrete layer. The shielding effect of the concrete layer is sub-dominant in comparison to the Earth's atmosphere shielding for the entire parameter space considered in this paper. }. A modified version of this Monte-Carlo simulation is used to derive the limits for DAMIC by considering the Earth's crust and the lead layer around DAMIC as the main shielding layers. The Monte-Carlo simulation presented here is an improved version of the DM{\scriptsize ATIS} code~\cite{code} which
\begin{itemize}
	\item \textit{Considers geometry of the overburden.} Modeling the Earth's atmosphere as a planar layer, as was done in~\cite{Emken2018} underestimates the number of capable DM particles by a factor as large as 2, depending on DM mass (see figure~\ref{Hybrid_to_Planar}). In appendix~\ref{atm_analytic}, we show that for large incident zenith angles ($|\cos\theta_i|\leq0.16$), the planar model is not justified. In the planar approximation, a DM particle with zenith angle $\theta\approx90\degree$, almost has no chance of reaching the detector independent of the value of the column-depth (see eq.~\eqref{chi_h_inv}). This leads to underestimation of the number of particles deflected outside of the Earth's atmosphere in the planar approximation and consequently underestimation of the number of capable DM particles.
	\item \textit{Models the Earth reflection effect.} This is especially important for the CSR experiment with lower mass reach (see appendix~\ref{earth_reflection_effect}).
	\item Models momentum-transfer and velocity-dependent DM-nucleon scatterings in inhomogeneous targets (see appendix~\ref{vel_generalization_of_DMATIS}).
\end{itemize}
 Assuming an isothermal spherical density profile, the DM velocity distribution in the Galactic rest frame, $f_G(\vec{v})$, is characterized by a Maxwellian distribution with a velocity dispersion of $v_0$ truncated at an escape velocity $v_{esc}$. Correspondingly, the velocity of DM particles in the Earth's frame, before entering the overburden, can be calculated by subtracting the velocity of the Earth in the Galactic rest frame, $\vec{v}_{E}$, from the DM velocity in the Galactic rest frame $\vec{v}_G$. In this work, we take the Earth's velocity $\vec{v}_{E}=(39.14,\,230.5,\,3.57) \,\rm{km\cdot s^{-1}}$, velocity dispersion $v_0=220\,\rm{km\cdot s^{-1}}$ and escape velocity $v_{esc}=584\,\rm{km\cdot s^{-1}}$~\cite{Erickcek2007}.
 
 \textit{Step 0}: The DM particle's initial speed and direction is sampled from the velocity distribution above the Earth's atmosphere. For a given DM mass, the column depth to the next interaction is sampled from the distribution given in eq.~\eqref{evP_tex}. The density is integrated along the initial velocity vector of the DM particle to find the position of the first scattering.  If its initial energy is smaller than the minimum required energy to trigger the CSR detector, we count this DM particle as one of the particles which does not give a signal. If its energy is above the minimum energy and the particle is already at the CSR altitude, $z_0\leq z_{det}=0$, it is counted as a capable DM particle. Otherwise, the DM particle enters the first scattering iteration. 
 
 \textit{Step 1:  Choose the target nucleus.} The probability that a given DM particle scatters off a nucleus of mass number $A$ is
 \beq
 \begin{split}
 	\label{eP(A)_vb}
 	P(A)&= \frac{n_A(r_i)\,\sigma_A(v_{i-1})}{\sum_A n_A(r_i)\,\sigma_A(v_{i-1})}\;,
 \end{split}
 \eeq
 where summation is over mass constituents of the shielding layer under consideration\footnote{ See appendix~\ref{earth_crust_model} for the Earth's crust constituents and appendix~\ref{atm_model} for the Earth's atmosphere constituents.}. $n_A(r_i)$ is the number density of nuclei of mass number $A$ at the current position of the DM particle. $\sigma_{A}(v_{i-1})$ is the DM-nucleus cross section, which is calculated using the velocity of the DM particle before scattering. 
 
 For milli-charged DM with electric charge $\epsilon\,e$ and velocity $v$, the total DM-nucleus cross section is (see eq.~\eqref{esigmaTAMilli})
 \beq\begin{split}
 	\label{esigmaAMilli_tex}
 	\sigma_A(v)= \, \frac{4\,\pi\,Z_A^2\,\epsilon^2\,\alpha^2}{v^2}\,\left( \frac{1}{m_e^2\,\alpha^2}-\frac{1}{4\,\mu_A^2\,v^2}\right)\;,
 \end{split}
 \eeq
 where $Z_A$ is charge number of nucleus of mass number $A$, $\alpha$ is the Standard Model fine structure constant, $m_e$ is the electron mass, and $\mu_A$ is the DM-nucleus reduced mass. For direct detection experiments under consideration in this paper, the minimal nuclear recoil energy $E_{\rm{nr}}^{\rm{th}}$ is much bigger than $E_{\rm{nr},A}^{\,\rm {screen}}\equiv(\alpha\,m_e)^2/2\,m_A$. This guaranties the positivity of the above expression.
 
For the contact interaction (heavy mediator) case, in the commonly-used Born approximation, the spin-independent DM-nucleus cross section is
 \beq\begin{split}
 	\label{e_sigmavAF2}
 	\sigma_{A}(v)&=\sigma_p(v) \left(\frac{\mu_A}{\mu_p} \right) ^2A^2\,  \;.
 \end{split}\eeq
where $\sigma_p(v)=\sigma_0\,v^n$ is the DM-nucleon cross section we consider in this paper.

\textit{Step 2: Choose the CM scattering angle, $\xi_i^{\rm{CM}}$, fixing the final energy.} The differential DM-nucleus cross section of a milli-charged DM is (see eq.\eqref{edsAdErMilli})
\beq\begin{split}
	\label{edsAdErMilli_tex}
	\frac{d\,\sigma_A}{d\,E_{\rm{nr}}}&=\,\frac{2\,\pi\,Z_A^2\,\epsilon^2\alpha^2}{m_A\,v^2\,E_{\rm{nr}}^2} \,F_A^2(E_{\rm{nr}}) \;,
\end{split}
\eeq
where the $m_A$ is the mass of nucleus. 

For milli-charged DM, just like in ordinary electromagnetism, forward-scattering is favored due to the $\propto E_{\rm{nr}}^{-2}$ dependence of the differential cross section in eq.~\eqref{edsAdErMilli_tex}. The differential cross section provides all the information needed to calculate the scattering angle distribution. The cumulative probability distribution of the recoil energy is
\beq\begin{split}
	\label{euMilli}
	U(E_{\rm{nr}})\equiv \frac{\int^{E_{\rm{nr}}}_{E_{\rm{nr}}^{\,\rm {screen}}}\frac{d\,\sigma_A}{d\,E_{\rm{nr}}}\,d\,E_{\rm{nr}}}{\int^{E_{\rm{nr}}^{\rm{max}}}_{E_{\rm{nr}}^{\,\rm {screen}}}\frac{d\,\sigma_A}{d\,E_{\rm{nr}}}\,d\,E_{\rm{nr}}}\;,
\end{split}
\eeq
where $E_{\rm{nr}}^{\,\rm{max}}=(2\,\mu_A\,v)^2/2\,m_A$ is the maximum nuclear recoil energy and $E_{\rm{nr},A}^{\,\rm {screen}}\equiv(\alpha\,m_e)^2/2\,m_A$ is the minimum nuclear recoil energy due to screening of Coulomb-like force by electrons at the atomic scale.
By noticing $\cos\,\xi^{\rm{CM}}=1-2\,E_{\rm{nr}}/E_{\rm{nr}}^{\rm{max}}$, we sample $U$ in [0,1] and solve for $E_{\rm{nr}}$ from eq.~\eqref{euMilli} to sample for $\cos\,\xi^{\rm{CM}}$. 

For the heavy-mediator case, the differential DM-nucleus cross section depends on the nuclear recoil energy only through the form factor
\beq\begin{split}
	\label{edsAdEr_tex}
	\frac{d\,\sigma_A}{d\,E_{\rm{nr}}}&=\,\frac{m_A\,\sigma_A(v)}{2\,\mu^2_A\,v^2} \,F_A^2(E_{\rm{nr}})  \;.
\end{split}
\eeq
The effect of the form factor can be neglected, i.e. $F_A(E_{\rm{nr}})\approx1$, for DM of mass $\lesssim$100 GeV scattering off the light nuclei in the Earth's atmosphere. In this case, the scattering is isotropic in the CM frame, thus $\cos\,\xi_i^{\rm{CM}}\in[-1,1]$ has a uniform distribution. 

Given the target element mass number $A$, the ratio of the DM particle's energy after scattering $i$ to its energy before scattering $i$ is:
 \beq
 \label{eEiEi-1_tex}
 \frac{E_{i}}{E_{i-1}}=1-\frac{4\,\mu^2_{A}}{m\,m_{A}}\left(\frac{1-\cos\,\xi_i^{\rm{CM}}}{2} \right)\;,
 \eeq
 where $\xi_i^{\rm{CM}}$ is the scattering angle in the CM frame. 
 
 \textit{Step 3: Choose the azimuthal scattering angle, $\varphi_i$, fixing the new direction.}  The scattering angle in the lab frame, $\xi_i$, and the scattering angle in the CM frame, $\xi_i^{\rm{CM}}$, are related by
 \beq
 \label{eXi_tex}
 tan\,\xi_i=\frac{sin\,\xi_i^{\rm{CM}}}{m/m_{A}+cos\,\xi_i^{\rm{CM}}}\;.
 \eeq
The zenith angle is calculated recursively using:
 \beq
 \label{eThetai_tex}
 cos\,\theta_i=\,cos\,\xi_i\,cos\,\theta_{i-1}-sin\,\xi_i\,sin\,\theta_{i-1}\,cos\,\varphi_i\;.
 \eeq
 
 Generate for $\varphi_i$ a random number in $[0,2\pi]$, to calculate the zenith angle after scattering $i$ using eqs.~\eqref{eXi_tex} and~\eqref{eThetai_tex}.
 \par
 \textit{Step 4: Find the position of the next interaction.} The probability of a DM particle scattering exactly once in the column-depth interval $[\chi, \chi+d\chi]$, off of a nucleus in a target with a mixture of constituents, is the product of the probability that the DM particle has not scattered in the column depth $\chi$ and the probability that it scatters in the distance $d\chi$ (see appendix~\ref{vel_generalization_of_DMATIS})
\beq
\label{evPdchi_tex}\begin{split}
	P(\chi,\chi_{eff})\,d\chi&=\frac{d\chi}{\chi_{eff}(v)}\,Q(\chi,\chi_{eff})\;,
\end{split}\eeq
where $\chi_{eff}(v_i)\equiv\left( \sum_A\frac{\sigma_A(v_i)\,f_A}{m_A}\right)^{-1} $ is the mean column depth for a mix of nuclei which is calculated using DM's current speed $v_i$. The column depth distribution is
\beq
\label{evP_tex}\begin{split}
	P(\chi,\chi_{eff})&=\frac{1}{\chi_{eff}(v)}\exp\left( -\frac{\chi(r_i)}{\chi_{eff}(v_i)}\right)\,,
\end{split}\eeq
 where $\chi(r_i)\equiv\int_0^{r_i}\rho_T(r')d\,r'$ is the column-depth accumulated as a function of distance $r_i$. Eq.~\eqref{evP_tex} is used to sample the column depth $\chi_s = -\chi_{eff}(v_i) \ln(1-ran_i)$, where $ran_i\in[0,1]$ is a random number representing the column depth's cumulative probability.
To determine the position of the next scattering $r_i$, for zenith angles $|\cos\theta_i|\leq0.16$, the mass density is numerically integrated along the current DM velocity vector until it reaches the sampled column depth $\chi_s$. For zenith angles $|\cos\theta_i|\geq0.16$, we use the planar model of the Earth's atmosphere to analytically find the position of the next scattering (see appendices~\ref{atm_analytic},~\ref{atm_model}).
 \par
 Steps 1 -- 4 are repeated until each DM particle is categorized as follows
 \begin{itemize}
 	\item If at any stage the altitude of a DM particle is more than 84.8 km \footnote{ The total over-head column depth at 84.8 km above the sea level is $\sim$0.00038\% of the total over-head column depth at sea level (see table~\ref{t_atm_chi}).} (positive direction of $\hat{z}$ taken to be upward with $z=0$ at the Earth surface), that particle is not tracked any more and is counted as one of the particles which scatters out of the Earth's atmosphere. 
 	\item If the energy of a DM particle becomes smaller than the minimum energy, $E\leq E_{min}$, before reaching the CSR detector, that DM particle is no longer tracked and is counted as one of the events with energy below the detector's threshold.
 	\item If a DM particle reaches the CSR detector's altitude, $z\leq0$, with potentially enough energy to trigger the detector, $E\geq E_{min}$, that particle is counted as a capable DM particle which can potentially trigger the detector. To account for the Earth reflection effect\footnote{ This effect is stronger for the lighter DM particles (see figure~\ref{Earth_reflection}).}, we continue to track trajectories of such particles in the Earth's crust. In each passage of a DM particle from the Earth's atmosphere to the Earth's crust and from the Earth's crust to the Earth's atmosphere, if the energy condition is satisfied i.e. $E\geq E_{min}$, a capable DM particle is added to the list of capable DM particles. Doing so, one particle can potentially trigger the detector more than once.
 \end{itemize}
 
 \subsection{Non-isotropic incident flux}
 The Monte-Carlo simulation method described in the previous section can be done for any specified arrival direction and velocity distribution. But an efficient way to treat a non-isotropic distribution, for instance to study the time dependence for a particular detector, is to perform the Monte-Carlo simulation for an isotropic distribution, and then re-weight the results appropriately. Here we do that for the DAMIC experiment, verifying that the cross section limit is nearly the same for the real velocity distribution as for an isotropic distribution.  
 
\begin{figure}[tbp]\centering
	\begin{subfigure}{.32\textwidth}
		\centering
		\includegraphics[width=\linewidth]{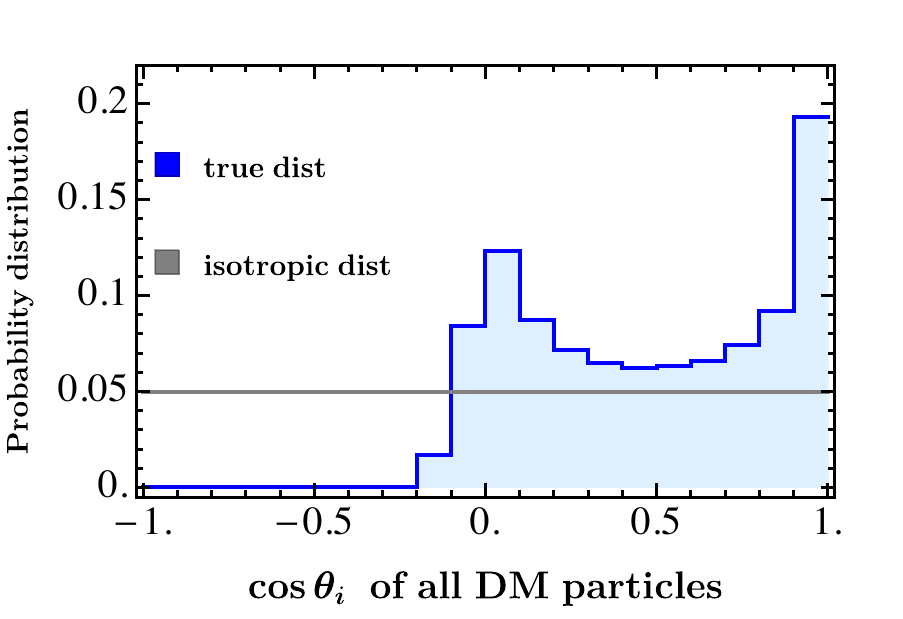}
		\caption{}
		\label{DAMICzenith_all}
	\end{subfigure}
	\begin{subfigure}{.32\textwidth}
		\centering
		\includegraphics[width=\linewidth]{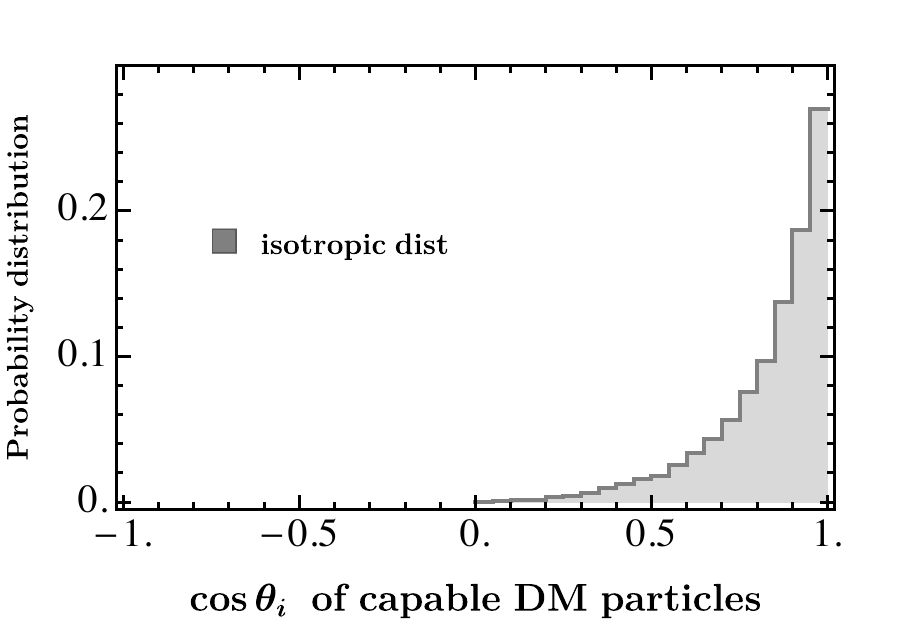}
		\caption{}
		\label{capableZenith_iso}
	\end{subfigure}
	\begin{subfigure}{.32\textwidth}
		\centering
		\includegraphics[width=\linewidth]{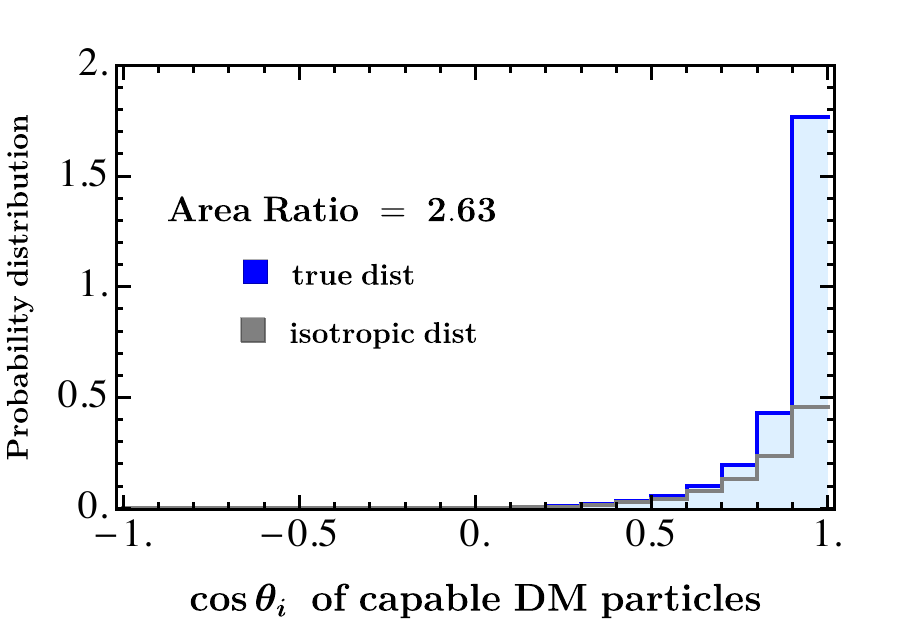}
		\caption{}
		\label{capableZenith}
	\end{subfigure}
	\caption{Various zenith-angle distributions relevant for DAMIC: (a) true incident particle zenith angle distribution for DAMIC exposure, (b) capable particles at the DAMIC detector, for an isotropic incident zenith angle distribution, (c) incident zenith angle distribution of capable particles at the DAMIC detector, for an isotropic incident distribution (grey, same as (b), shown for comparison) and for the actual incident zenith angle distribution (blue). }
	\label{DAMIC_Zenith}
\end{figure}

The zenith-angle distribution of incident DM particles at DAMIC, averaged over the 11 months of its run, is shown in figure~\ref{DAMICzenith_all}.
 In~\cite{method} we analyzed the properties of capable particles reaching the DAMIC depth. Figure 11a of \cite{method}, reproduced here as figure~\ref{capableZenith_iso}, shows the distribution of  incident zenith angles of \textit{capable} particles for a DM mass of 1.7 GeV. As can be seen, the requirement of arriving at the depth of DAMIC without too much energy loss, strongly biases the distribution of zenith angles relative to isotropic, favoring vertical particles. The distribution in figure~\ref{capableZenith_iso}, normalized to unity, is the relative probability distribution for a given zenith angle of incidence to produce a capable particle. The product of the distributions (a) and (b) gives the actual distribution of arrival directions of capable particles at the DAMIC detector. The area ratio is 2.63, meaning that for a given DM cross section, DAMIC should see 2.63 times more events for the real flux geometry than for an isotropic arrival distribution. This will improve its cross-section reach relative to what we report, but only by $\mathcal{O}(1$\%).  
 
The date and time of the CRS exposure are not reported in the papers, so we cannot do the same analysis for CSR because we cannot produce the CSR analog of figure~\ref{DAMICzenith_all}. But the DAMIC example demonstrates that refining the zenith angle distribution for CSR should not appreciably change the CSR reach reported below.    
  
 \par
\section{Multiple scatterings in the XQC detector}\label{multiple_scatterings}
As pointed out by Erickcek et al.~\cite{Erickcek2007}, considering multiple scatterings in the XQC detector extends the mass reach of the XQC detector, down to 10 MeV for thermalization efficiency $\epsilon_{\rm{th}}=1$. In the limit of single scattering in the XQC detector,  Eq.~\eqref{edNdEr_gen} can be used to calculate the expected spectrum for DM masses $\geq$ 225 MeV for $\epsilon_{\rm{th}}=1$, masses $\geq$ 1 GeV for $\epsilon_{\rm{th}}=0.1$, and masses $\geq$ 2 GeV for $\epsilon_{\rm{th}}=$ 0.02.  But, this equation cannot be used in the case of multiple scatterings.   In the analysis reported below, we therefore consider multiple scatterings in the XQC detector, to determine the mass reach for other thermalization efficiency factors. This enables us to reach down to 420 MeV for $\epsilon_{\rm{th}}=$ 0.02, the most conservative XQC thermalization efficiency factor that we consider in this work.  In this section, we review the basics of our Monte-Carlo simulation\footnote{ For XQC, we use the full 3D velocity distribution of incoming DM particles and take into account the anisotropic zenith angle distribution at the time and location of the XQC detection.} that calculates the total recoil energy deposited in the XQC detector in the multiple-scattering regime.

The aluminum body of the rocket carrying XQC produces an overburden of $\approx10\,\rm{g\cdot cm^{-2}}$\cite{McCammon2002, ErickcekThesis}, corresponding to 3.7 cm aluminum. In the limit of multiple scatterings, i.e. $\lambda_{eff}\lesssim 0.2$ cm (where 0.2 cm is the largest edge of the XQC calorimeter), the body of the rocket shields the XQC detector and only DM particles coming through the 1 steradian opening angle centered on the normal to the detector, $(l,b)=(90\degree, 60\degree)$, would reach the detector. Therefore, a DM particle has to enter the XQC detector through its top face with surface area $S=10^{-2}\rm{cm^{2}}$. 
\begin{figure}[tbp]\centering
	\includegraphics[width=1\textwidth]{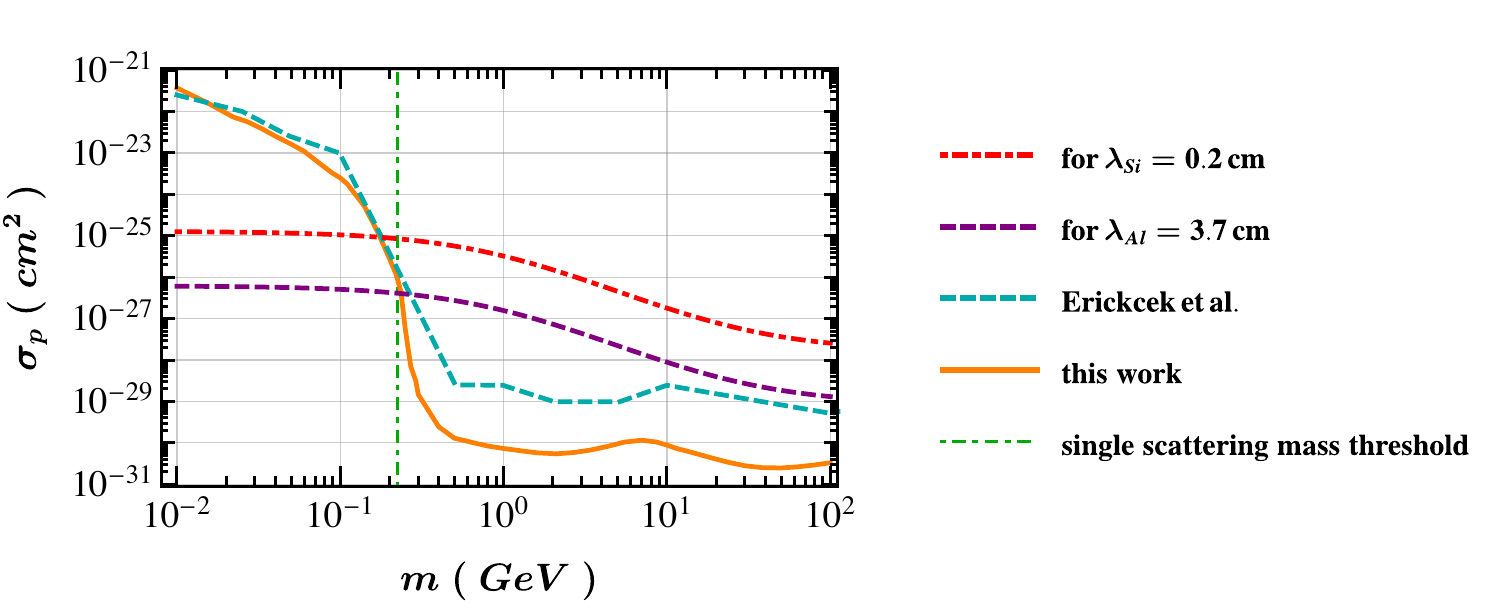}
	\caption{\label{XQC_limits_0}The XQC 90\% CL lower reach on velocity-independent DM-proton cross section, assuming thermalization efficiency $\epsilon_{\rm{th}}=1$. The orange solid line shows the result of this work. The dashed cyan line is read from the Erickcek et al. paper~\cite{Erickcek2007} where the body of the rocket was assumed to shield the XQC detector for all DM masses. The purple dashed (red dash-dotted) line show the DM-proton cross section required for a 3.7 cm (0.2 cm) interaction length in aluminum (silicon). The green dash-dotted line shows the single scattering mass threshold for XQC. See the text for a detailed discussion.}
\end{figure}
In the multiple scatterings case, we model each XQC calorimeter as a slab, made of silicon\footnote{ DM particles with masses $\leq$2 GeV lose most of their energy to silicon nuclei and not to nuclei in the HgTe layer of the detector due to negligible fractional energy loss to heavier nuclei (see eq.~\eqref{eEiEi-1_tex})}, with dimensions $(d_x, \,d_y,\,d_z)=(0.2,\,0.05,\,1.6\times10^{-3})$ cm. We divide the surface of each calorimeter into 100 squares with equal surface area $\delta S=10^{-4}\rm{cm^{2}}$. Using this mesh enables us to take into account the geometry of the detector in the multiple scatterings case. 

Then, we inject an incoming DM particle randomly to one of the differential surface areas and follow its trajectory in the XQC detector until
\begin{itemize}
	\item Its energy falls below 1 eV, or
	\item Its new position is outside the boundaries of the XQC detector. 
\end{itemize}
Steps of the Monte-Carlo simulation for this calculation are the same as what is described in section~\ref{MC_description} except that
\begin{itemize}
	\item We track the total amount of deposited energy in the XQC detector.
	\item We track the position of the DM particle in three dimension to account for the geometry of the XQC detector.
\end{itemize}

Figure~\ref{XQC_limits_0} shows the XQC 90\% CL lower reach on the velocity-independent DM-proton cross section, assuming thermalization efficiency $\epsilon_{\rm{th}}=1$ in order to compare to the earlier work of Erickcek et al.~\cite{Erickcek2007}. The orange solid line shows the result of our analysis. The dashed cyan line is read from the exclusion plot in~\cite{Erickcek2007}, where the body of the rocket was assumed to shield the XQC detector for all DM masses. The purple dashed line shows the DM-proton cross section required for a 3.7 cm interaction length in aluminum (the thickness of the body of the rocket carrying XQC). Therefore, as was shown in~\cite{bounds}, since the actual cross section limit is smaller than this value the rocket body does not shield the XQC detector in the DM mass range 300 MeV -- 100GeV. The red dot-dashed line shows the DM-proton cross section required for a 0.2 cm (the largest edge of the XQC calorimeter) interaction length in silicon. For cross sections larger than this, multiple scatterings become relevant. Considering multiple scatterings in the XQC detector extends the mass reach of XQC from 225 MeV (minimum DM mass that can be constrained in the single scattering limit) to 10 MeV. For DM masses $\lesssim$ 225 MeV, our result is in a reasonable agreement with the result from Erickcek et al.~\cite{Erickcek2007} for velocity independent cross section and $\epsilon_{\rm{th}}=1$.  The results presented in the next section take into account multiple scattering where relevant.

\section{Results}\label{result_vb}
\begin{figure}[tbp]\centering
	\begin{subfigure}{.495\textwidth}
		\centering
		\includegraphics[width=\linewidth]{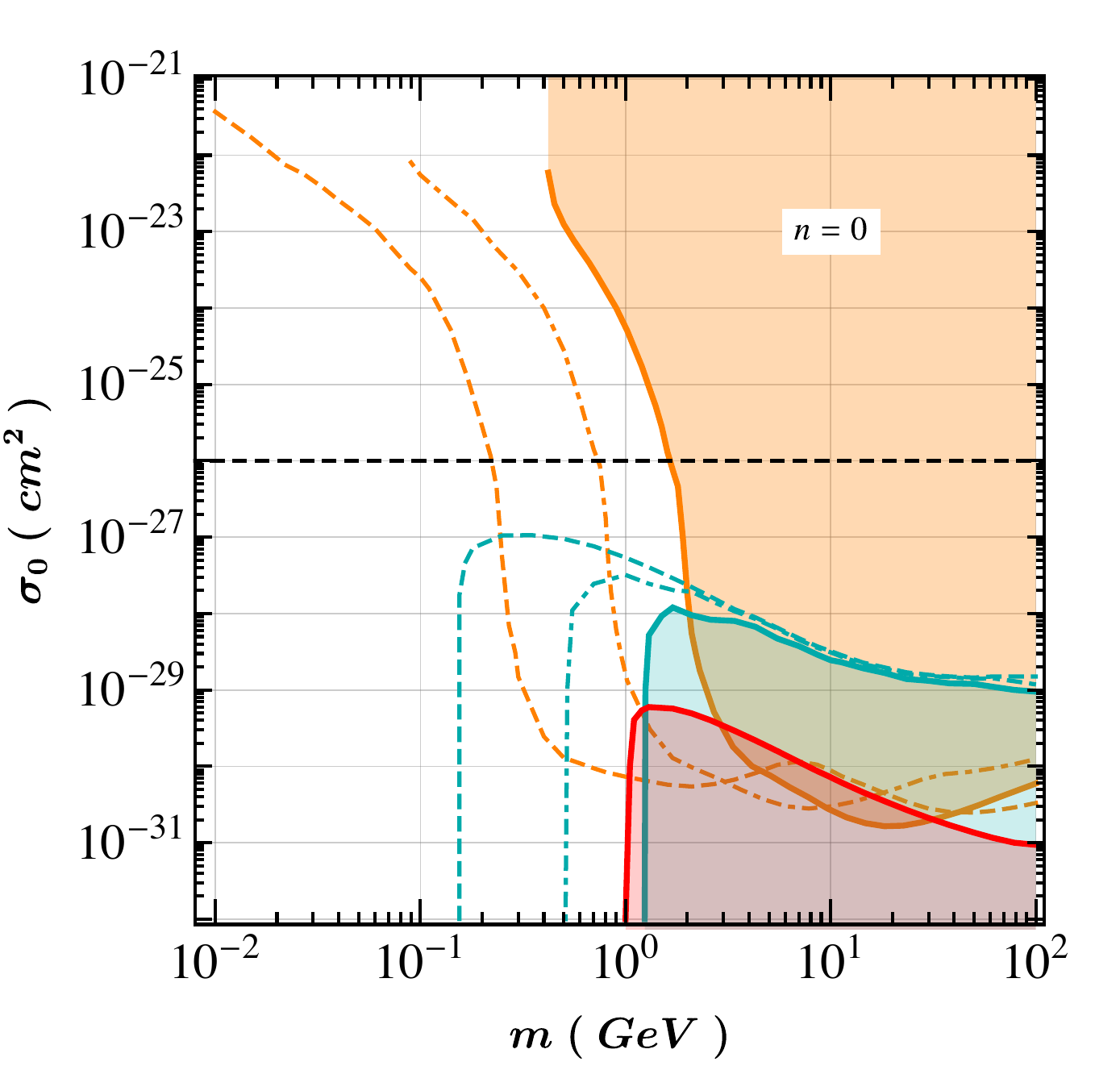}
		\caption{}
		\label{XND0}
	\end{subfigure}
	\begin{subfigure}{.495\textwidth}
		\centering
		\includegraphics[width=\linewidth]{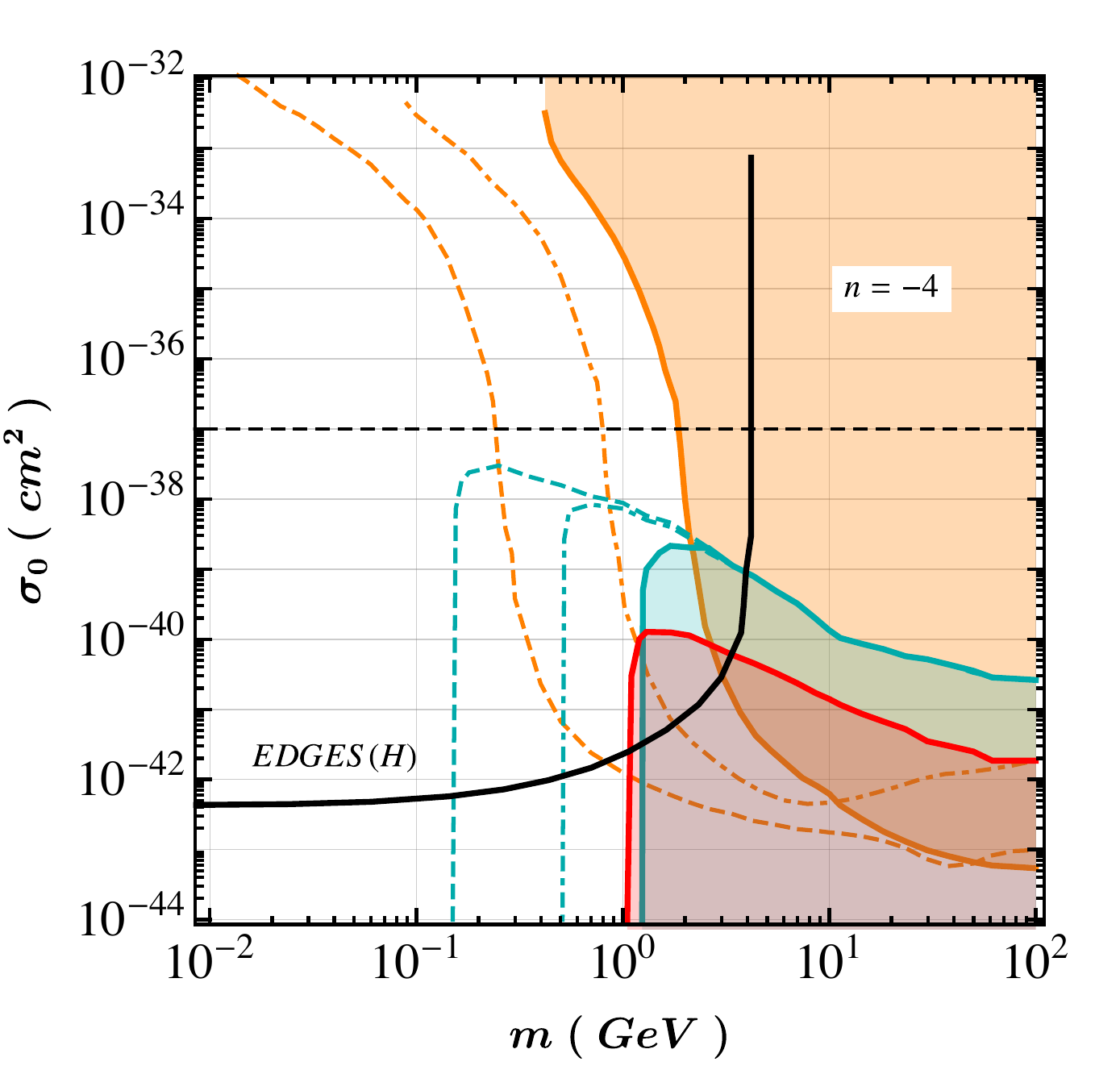}
		\caption{}
		\label{XNDn4}
	\end{subfigure}
	\caption{The 90\% CL DM-nucleon excluded regions bounded above by CSR (CRESST 2017 surface run shown in cyan solid line) and below by XQC (shown in orange solid line) for a DM with (a) velocity-independent and (b) power-law velocity-dependence $\sigma=\sigma_0\,v^{-4}$ cross section, and mass in the range 10 MeV -- 100 GeV. The solid lines are for a conservative estimate of the thermalization efficiency $\epsilon_{\rm{th}}=0.02$. Dashed (dash-dotted) lines show the CSR and XQC limits calculated for thermalization efficiency $\epsilon_{\rm{th}}=1\, (0.1)$. Limits from DAMIC (red lines), which were the strongest lower bounds before CSR, are shown for comparison, for $\epsilon_{\rm{th}}=1$. The cross section required to fit the EDGES signal for DM-hydrogen interactions (black solid line) is taken from~\cite{Barkana:2018qrx}. Dashed thin black lines show $\sigma_0$ values above which multiple scatterings in the XQC detector become relevant. }
	\label{XND1}
\end{figure}
\begin{figure}[tbp]\centering
	\begin{subfigure}{.495\textwidth}
	\centering
	\includegraphics[width=\linewidth]{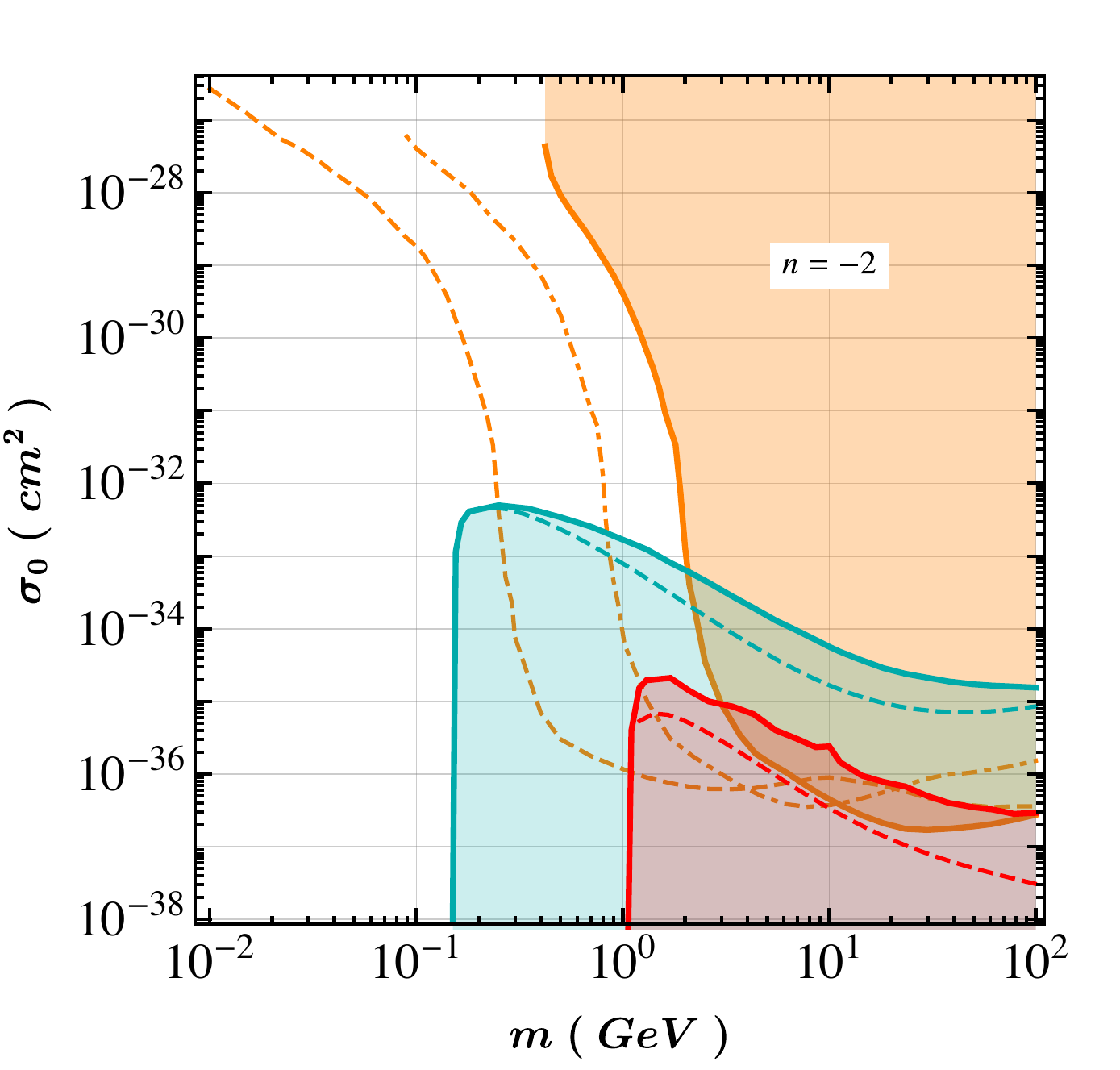}
	\caption{}
	\label{XNDn2}
	\end{subfigure}\hspace{.01cm}
	\begin{subfigure}{.495\textwidth}
	\centering
	\includegraphics[width=\linewidth]{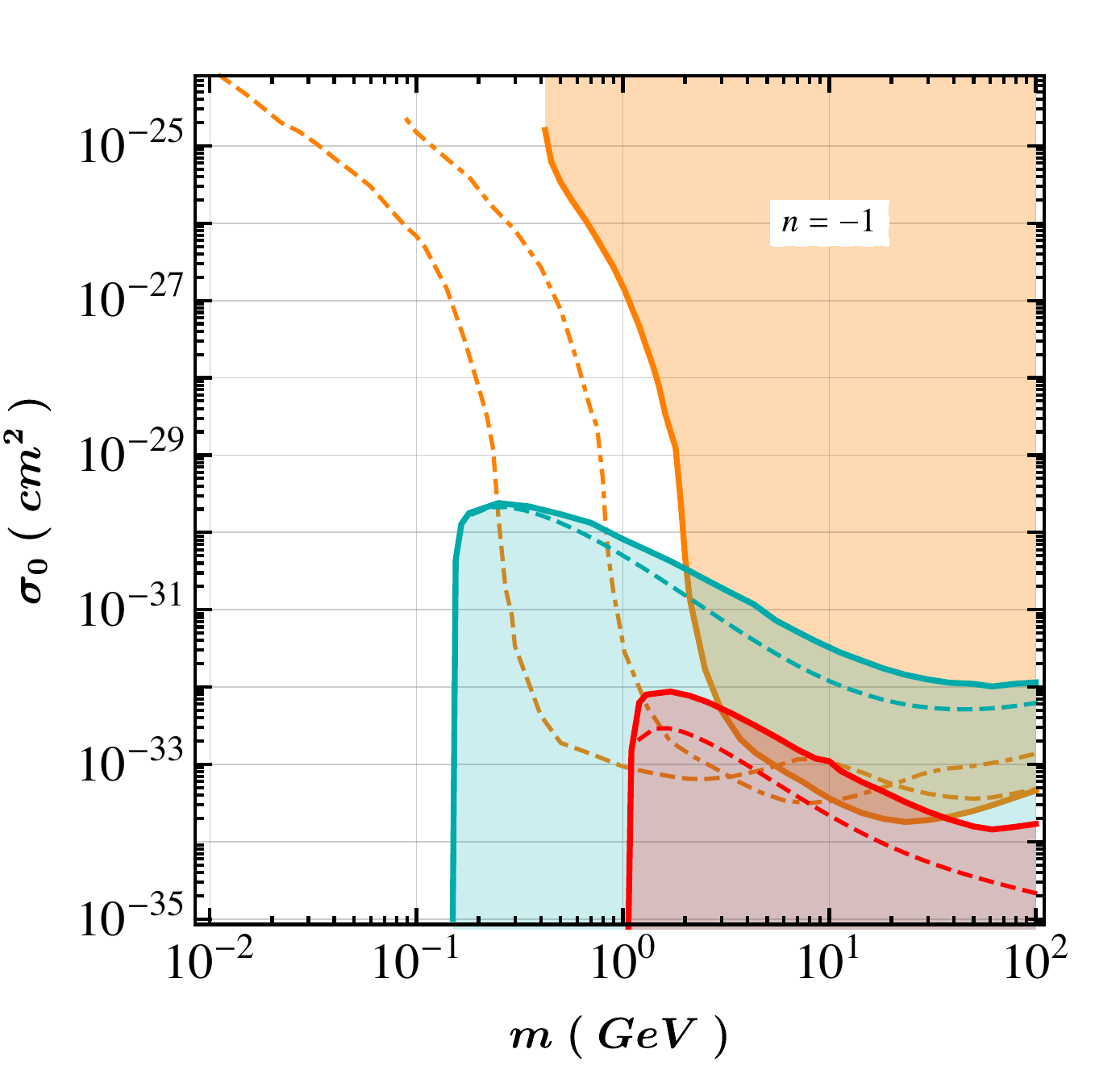}
	\caption{}
	\label{XNDn1}
	\end{subfigure}
	\\
	\begin{subfigure}{.495\textwidth}
		\centering
		\includegraphics[width=\linewidth]{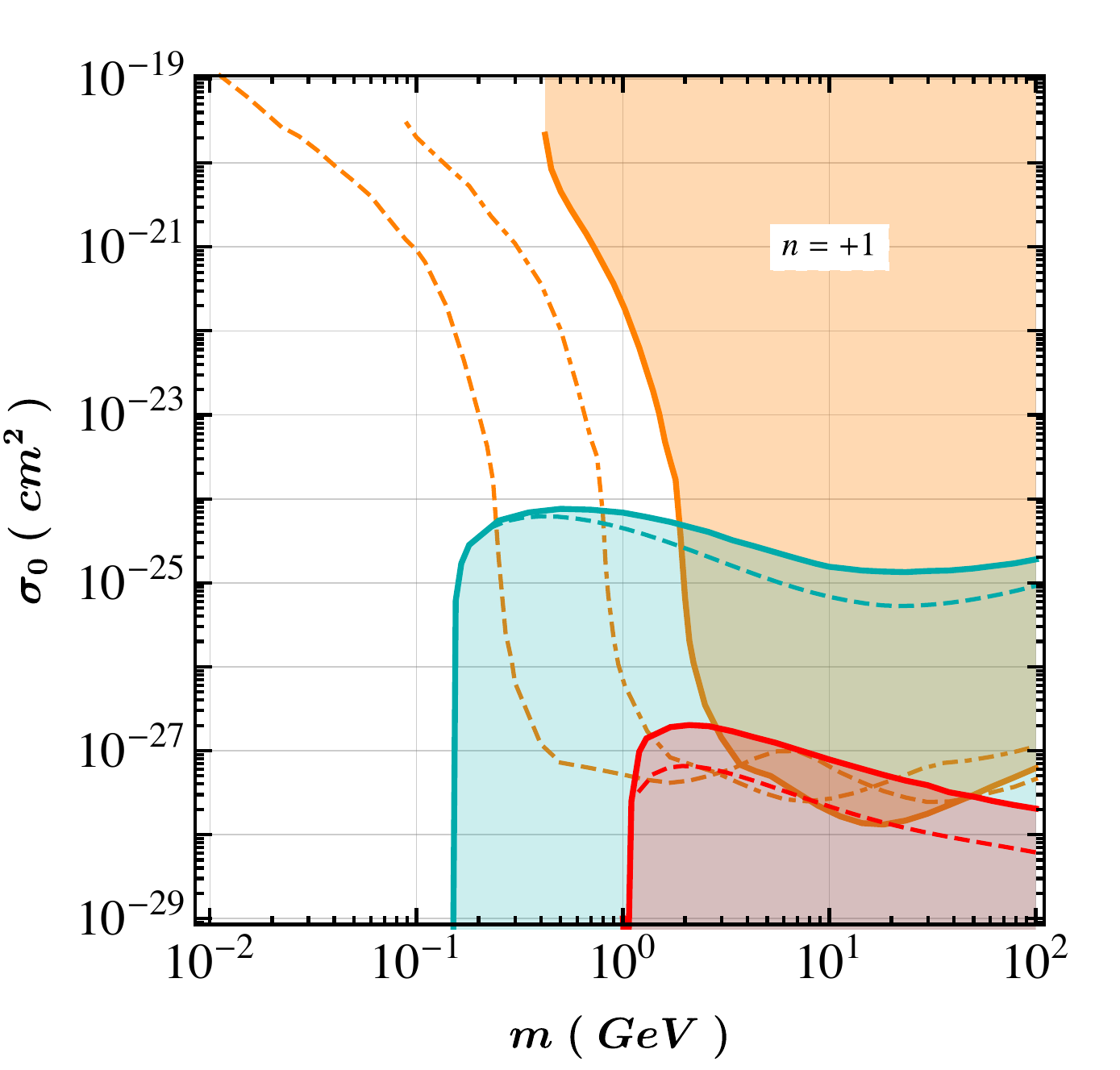}
		\caption{}
		\label{XNDp1}
	\end{subfigure}\hspace{.01cm}
	\begin{subfigure}{.495\textwidth}
		\centering
		\includegraphics[width=\linewidth]{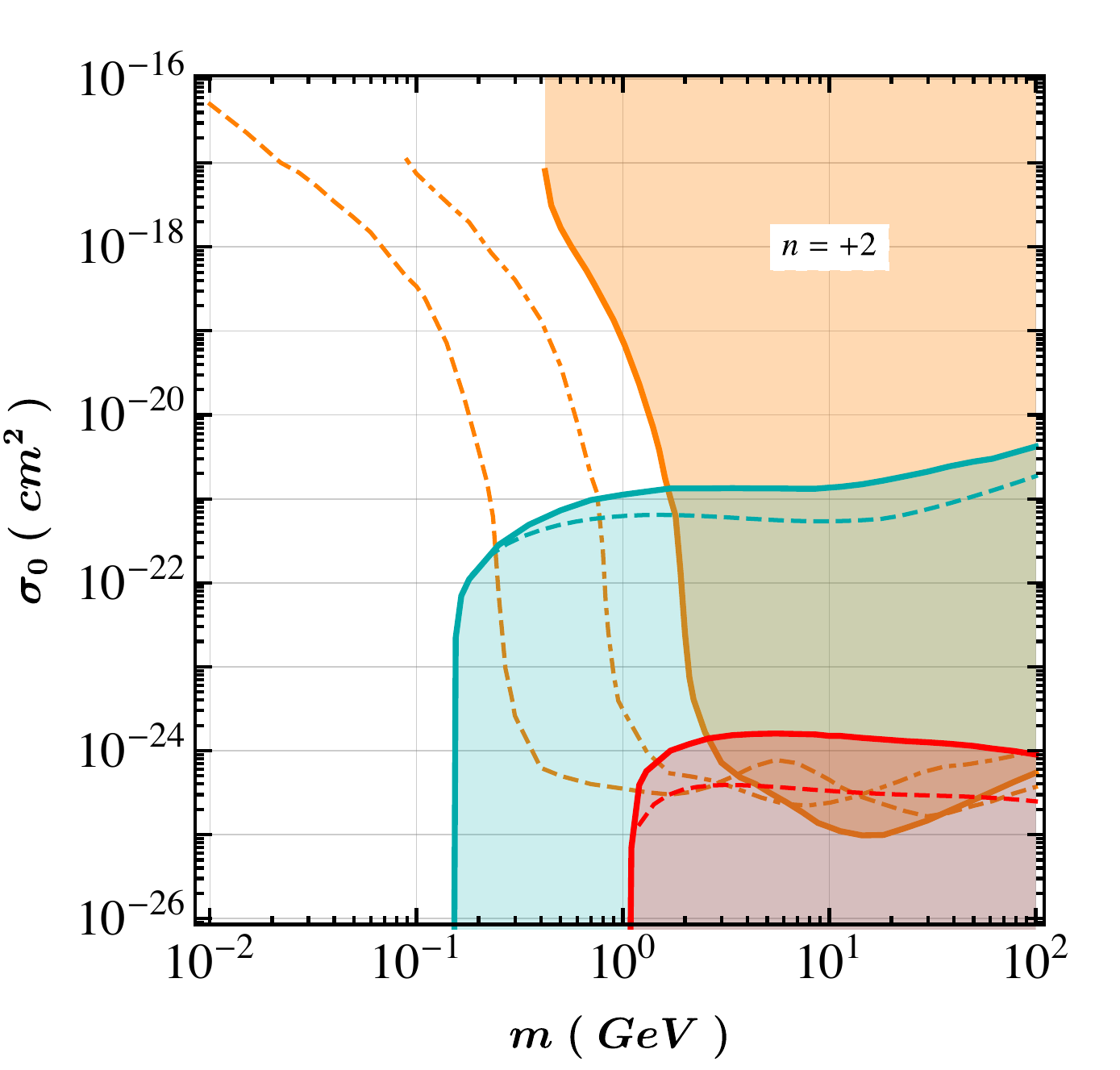}
		\caption{}
		\label{XNDp2}
	\end{subfigure}
	\caption{\label{XND2} Similar to figure~\ref{XND1}, but for DM-nucleon cross section with power-law velocity-dependence of the form $\sigma=\sigma_0\,v^n$ with $n=$\{-2,-1,1,2\}. Dashed cyan (red) line shows the CSR (DAMIC) limits calculated using the SGED approximation described in appendix~\ref{SGEDvel}.}
\end{figure}

Figures~\ref{XND1} and ~\ref{XND2} summarize our 90\% CL cross section bounds for DM masses of 10 MeV -- 100 GeV, using the XQC, DAMIC, and CRESST 2017 surface run (CSR) energy-deposit spectra, assuming power-law velocity-dependence of the form $\sigma=\sigma_0\,v^n$ with $n=$\{-4,-2,-1,0,1,2\} and using the Born approximation relation between cross sections for different nuclei. 

The XQC lower reach on $\sigma_0$ is calculated for a thermalization efficiency factor $\epsilon_{\rm{th}}=0.02$ (orange solid lines), for $\epsilon_{\rm{th}}=0.1$ (orange dash-dotted lines), and for full thermalization efficiency, i.e. $\epsilon_{\rm{th}}=1$ (orange dashed lines). As discussed in the appendix~\ref{ap:effth}, the thermalization efficiency may be much lower than $1$ as had been previously uncritically assumed. Clearly, measuring the thermalization efficiency is essential to determine what portion of the parameter space is allowed or excluded. 

We extended the mass reach of XQC to lower masses --- from 2 GeV to 420 MeV for $\epsilon_{\rm{th}}=0.02$ and from 225 MeV to 10 MeV for $\epsilon_{\rm{th}}=1$ --- by allowing for multiple scatterings of an individual DM particle in the XQC detector. In the mass range where single scattering gives the strongest limits, shown as a dashed thin black line in figure~\ref{XND1}, the 90\% CL XQC limits for a power-law index $n_1$ can be calculated by multiplying the limits for a different power-law index $n_2$ by $\left( \frac{V_{\rm{XQC}}}{c}\right) ^{n_1-n_2}$ due to the linear cross section dependence of the expected spectrum\footnote{ $V_{\rm{XQC}}$ is the mean velocity of DM particles above the XQC threshold.}.
 
The CSR experiment with threshold nuclear recoil energy $E_{\rm{nr}}^{\rm{th}}=19.7$ eV (assuming full thermalization efficiency) extends the 1 GeV mass reach of the hadronically-interacting DM exclusion region obtained in~\cite{bounds} using DAMIC with $E_{\rm{nr}}^{\rm{th}}=550$ eV, down to 150 MeV. Owing to its minimal shielding in comparison to underground experiments, the CSR cross section bounds are stronger by a factor of 10 -- 3000 in comparison to the DAMIC bounds, depending on the DM mass and the DM-nucleon cross section velocity-dependence index $n$. The bounds that we find using Monte-Carlo simulation are stronger by a up to a factor 4.3 for dark matter mass in the 400 MeV -- 100 GeV range, in comparison to the result of the SGED approximation (see appendix~\ref{SGEDvel}). For DM masses below 400 MeV, due to the Earth reflection effect described in appendix~\ref{earth_reflection_effect}, the bounds that we find using the Monte-Carlo simulation are approximately equal to the result of the SGED approximation. 

\begin{figure}[tbp]\centering
	\includegraphics[width=0.8\textwidth]{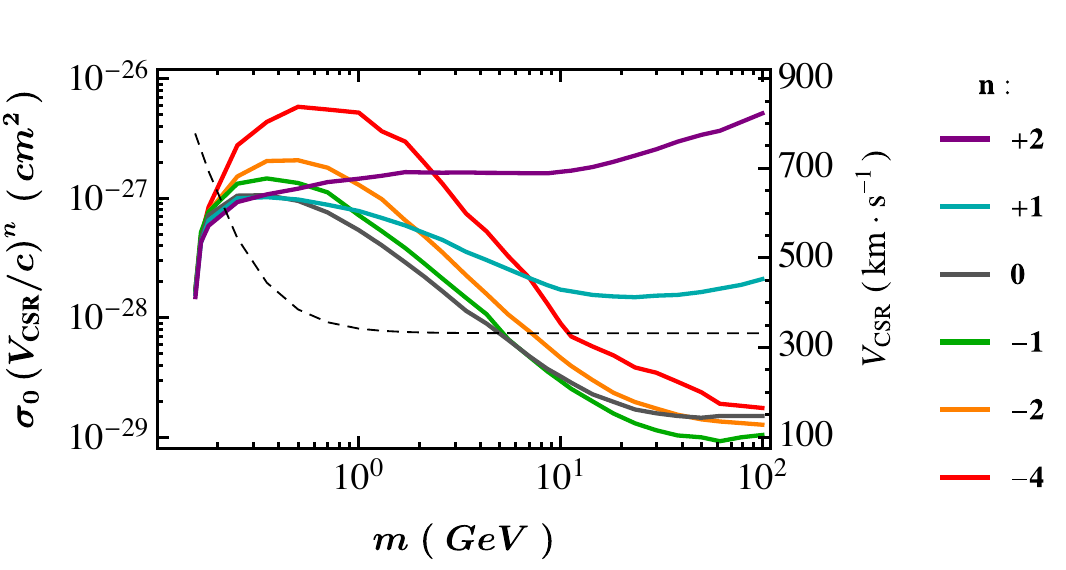}
	\caption{\label{NU_constraining_power} The 90\% CL upper reach of CSR (CRESST 2017 surface run) in the $m-\sigma_0\,(v/c)^n$ parameter space assuming full thermalization efficiency, i.e. $\epsilon_{\rm{th}}=1$. Here, we evaluate the cross section limits at $v=V_{\rm{CSR}}$, the mean velocity of DM particles before entering the Earth's atmosphere which are above the CSR threshold. This plot shows the relative sensitivity of CSR as a function of DM mass, for different assumed velocity-dependencies. $V_{\rm{CSR}}$ as a function of the DM mass is shown in the thin black line, with coordinate on the right.}
\end{figure}

Figure~\ref{NU_constraining_power} shows the 90\% CL upper reach of CSR in the $m-\sigma_0\,(v/c)^n$ parameter space. To facilitate the comparison, we introduce $v=V_{\rm{CSR}}$, the mean velocity of DM particles before entering the Earth's atmosphere which are above the CSR threshold. We evaluate the cross section for velocity $V_{\rm{CSR}}$, to introduce a metric to quantify the constraining-capability of CSR as a function of the DM mass and the power-law index of the DM-nucleon cross section. For DM masses $\geq3$ GeV, the CSR limits are stronger for positive power-law indices in comparison to negative power-law indices.  The velocity of DM particles decreases as these particles travel through an overburden. This causes the DM-nucleon cross section with positive (negative) power-law indices to decrease (increase) during the passage of DM particle through an overburden. Therefore, DM particles with positive velocity dependences can accommodate larger cross sections. 

Figure~\ref{Miili_plot} shows the 90\% CL constraints for milli-charged DM on charge, $\epsilon$, assuming the milli-charged particle constitutes (a) 100\% (b) 1\% of the DM density in the Earth's rest frame, as a function of the DM mass. The various lines are as follows:
\begin{figure}[tbp]\centering
	\begin{subfigure}{.495\textwidth}
		\centering
		\includegraphics[width=\linewidth]{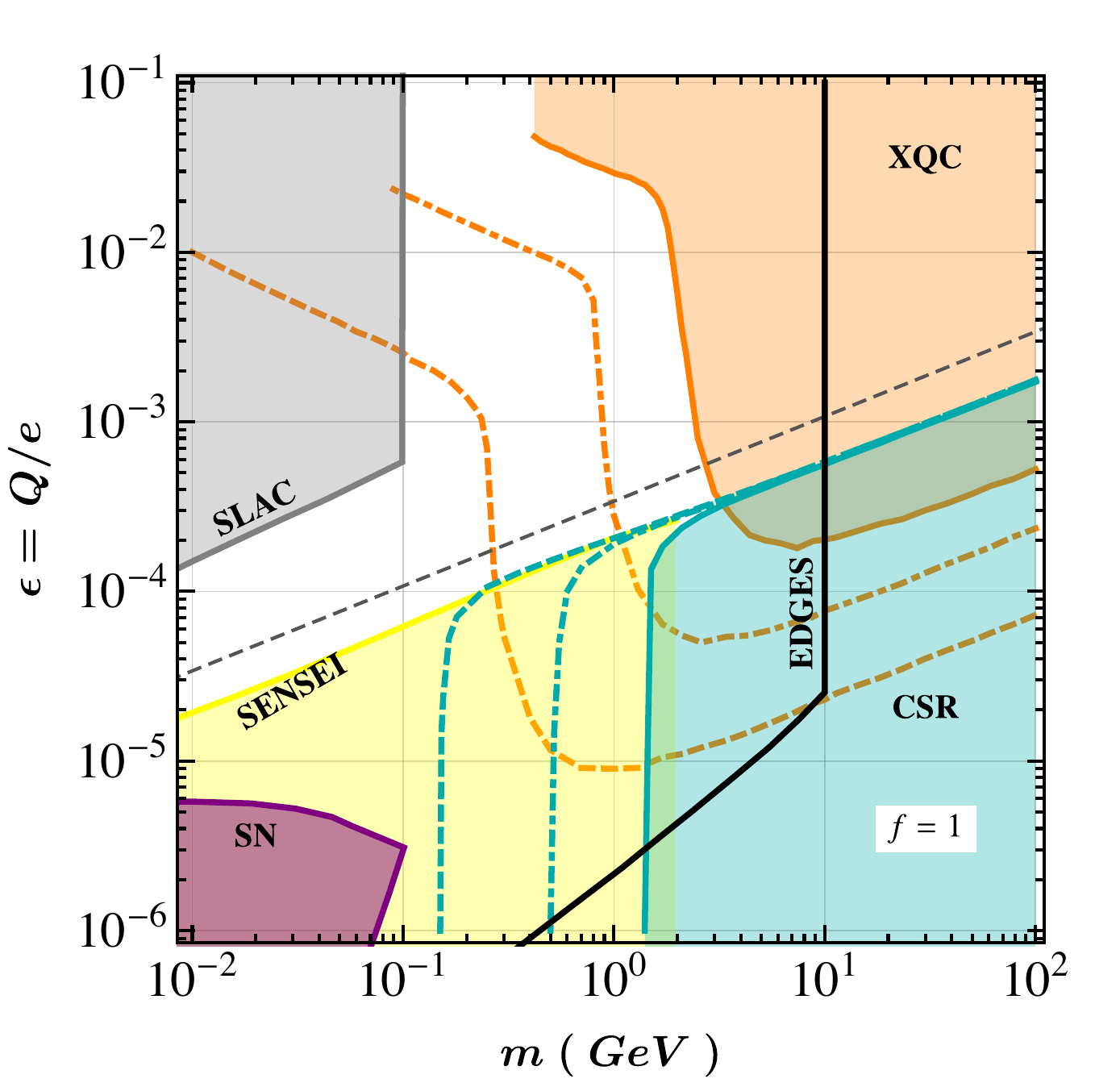}
		\caption{}
		\label{Miili_epsilon100}
	\end{subfigure}\hspace{.01cm}
	\begin{subfigure}{.495\textwidth}
		\centering
		\includegraphics[width=\linewidth]{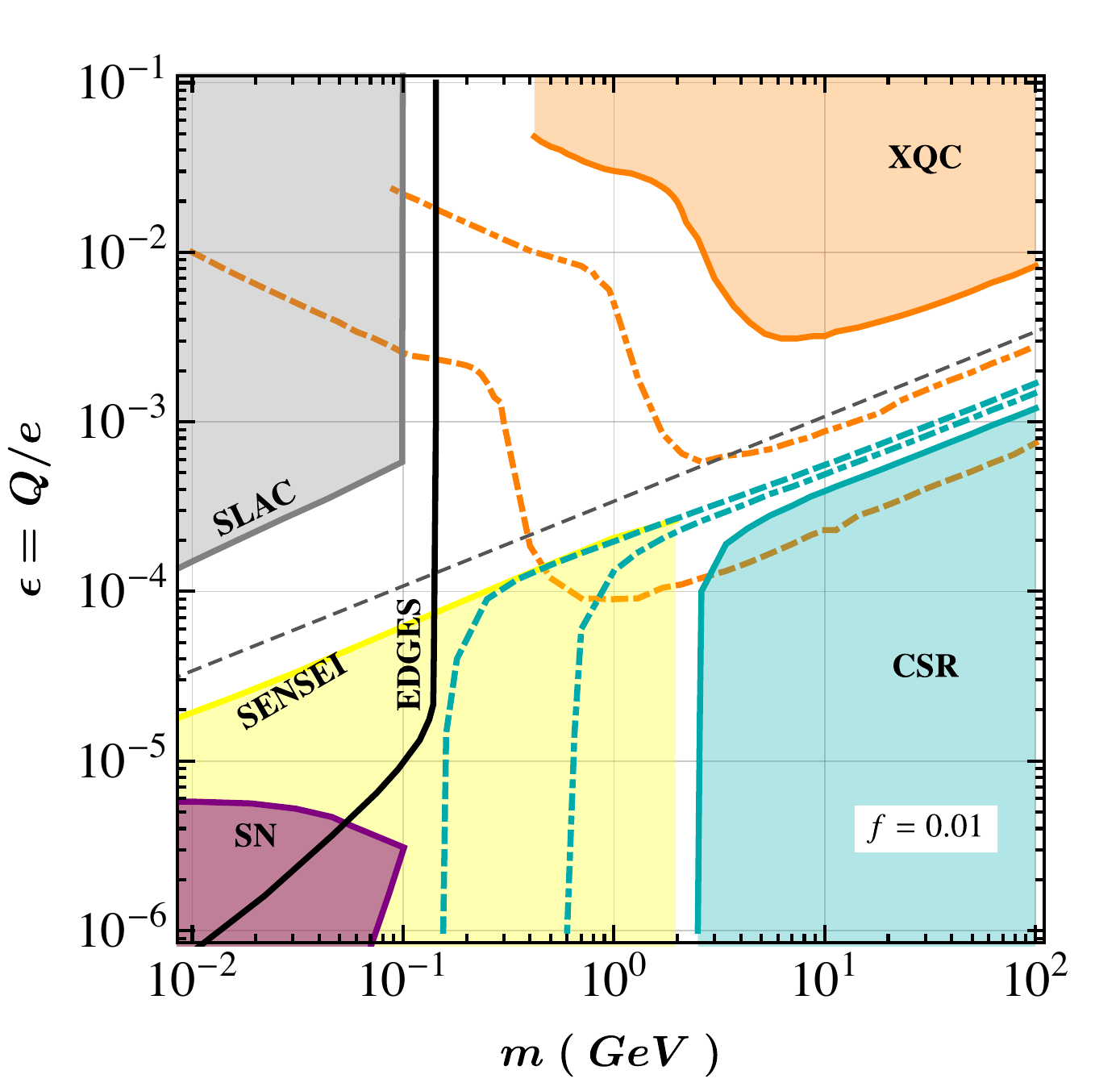}
		\caption{}
		\label{Miili_epsilon1}
	\end{subfigure}
	\caption{\label{Miili_plot} Excluded regions for the milli-charged DM model (90\% CL) on charge, $\epsilon$, assuming the milli-charged particle constitutes (a) 100\% (b) 1\% of the DM density in the Earth's rest frame, as a function of the DM mass. The cyan solid [dash-dotted, and dashed] line shows the upper reach of CSR (CRESST 2017 surface run) for thermalization efficiency $\epsilon_{\rm{th}}=0.02\;[0.1,$ and $1]$. The orange solid [dash-dotted, and dashed] line shows the lower reach of XQC for thermalization efficiency $\epsilon_{\rm{th}}=0.02\;[0.1,$ and $1]$. The black solid lines indicate the minimal charge needed to explain the EDGES measurement~\cite{Barkana:2018qrx}. The excluded regions from cooling of the supernova SN 1987A~\cite{Chang:2018rso} is shown in purple, from SLAC millicharge experiment~\cite{Prinz:1998ua} is shown in gray, and from SENSEI~\cite{Crisler:2018gci} is shown in yellow. The thin gray dashed line shows the maximum charge of a milli-charged DM particle that can be evacuated by supernova explosion according to~\cite{McDermott:2010pa}.}
\end{figure}
\begin{itemize}
	\item The cyan solid [dash-dotted, and dashed] line shows the upper reach of CSR for thermalization efficiency $\epsilon_{\rm{th}}=0.02\;[0.1,$ and $1]$. For DM masses $\gtrsim$ 400 MeV, the upper reach of CSR for milli-charged particles constituting 100\% of the DM density, is only 5\% -- 10\% larger in comparison to the case where milli-charged particles constituting 1\% of the DM density in the Earth's rest frame. 
	
	The upper reach of CSR is quite insensitive to the DM density before entering the Earth's atmosphere. This is due to the rapid increase in the flux of capable DM particles at the detector, with only a 5\% -- 10\% decrease of the DM-nucleon cross section near the limiting value. This is a common feature of limits for which energy-loss in an overburden determines the constraining-capability of a direct detection experiment~\cite{bounds, method, Emken2018}. 
	
	\item The orange solid [dash-dotted, and dashed] line in figure~\ref{Miili_plot} shows the lower reach of XQC for thermalization efficiency $\epsilon_{\rm{th}}=0.02\;[0.1,$ and $1]$.
\end{itemize}
Generally, when the effect of shielding materials around the detector and multiple scatterings in the detector can be ignored, the lower reach of direct detection experiments on the DM-nucleon cross section is inversely proportional to the DM density in the Earth's rest frame (and correspondingly the minimum excluded value of charge is $\propto1/\sqrt{\rho}$). This is the case for thermalization efficiency $\epsilon_{\rm{th}}=1$ and DM masses $\gtrsim$ 300 MeV, so the limits for when milli-charged particle constitutes 1\% of the DM density in the Earth's rest frame are weaker by a factor of $\sqrt{100}=10$, as seen by comparing the orange dashed lines for $\gtrsim$ 300 MeV in figures~\ref{Miili_epsilon100} and~\ref{Miili_epsilon1}. However, this is not the case for thermalization efficiency $\epsilon_{\rm{th}}=0.02$ and DM masses $\gtrsim$ 300 MeV. For this case, the limits for when milli-charged particle constitutes 1\% of DM density in the Earth's rest frame are weaker by a factor 14 -- 16 (compared with the naive factor of 10) due to the substantial shielding by the aluminum body of the rocket carrying XQC, as seen by comparing the orange solid lines for $\gtrsim$ 300 MeV in figures~\ref{Miili_epsilon100} and~\ref{Miili_epsilon1}.

In the regime where multiple scatterings in the XQC detector gives the strongest limits, i.e. DM masses $\lesssim$ 225 MeV for thermalization efficiency $\epsilon_{\rm{th}}=1$, DM masses $\lesssim$ 1 GeV for thermalization efficiency $\epsilon_{\rm{th}}=0.1$, and DM masses $\lesssim$ 2 GeV for thermalization efficiency $\epsilon_{\rm{th}}=0.02$, the limits are similar for different DM densities due to the strong cross section dependence of the energy-loss mechanism in this regime.

Unlike the constraints from the effective number of relativistic particles at CMB and at BBN~\cite{Vogel:2013raa,Brust:2013xpv, Foot:2014uba}, direct detection limits on milli-charged DM don't rely on the existence of a dark photon. The limits from XQC and CSR that we considered in this work rely on the DM-proton elastic scattering. These two experiments, with significantly smaller shielding overburdens in comparison to underground experiments, enable us to constrain much of the parameter space of milli-charged DM for charge values $10^{-6}$ -- $10^{-1}$ and DM masses 10 MeV -- 100 GeV.  

As we were completing this manuscript, ref.~\cite{Crisler:2018gci} appeared which finds the limits using the results of the new SENSEI Surface Run experiment. For both the limits from CSR (this work) and SENSEI Surface Run~\cite{Crisler:2018gci}, based on the result of ref~\cite{Kouvaris2014}, the electronic energy-loss in the Earth's atmosphere is neglected and the limits are calculated by considering only the nuclear energy-loss. As both CSR and SENSEI Surface Run were on the Earth's surface and thus have the same over-head column-depth, the upper-reach of CSR overlaps with that of SENSEI in the 150 MeV -- 2 GeV mass range.  The SENSEI limits complement the CSR limits by extending the mass reach down to $\sim$500 keV.  Although SENSEI makes powerful contributions to limits on milli-charged DM, it is not relevant for hadronically interacting DM, since it is only sensitive to DM-electron interactions.

Note that energy-loss by interactions with electrons in the overburden has not been included in our milli-charge analysis.  Doing so will mildly reduce the sensitivity of both CSR and SENSEI surface runs.   However the lower-reach limits for XQC that we present here will not be affected by inclusion of electronic energy-loss because all forms of energy-loss are negligible given the small overburden.  This makes the XQC limits the most stringent limits from direct detection experiments, for the large-$\epsilon$ part of the parameter space of milli-charged DM.  

\section{Conclusion}
In this paper, we derive for the first time the constraints on interacting Dark Matter with velocity-dependent DM-nucleon cross section using the 
XQC, DAMIC, and CRESST 2017 Surface Run (CSR) experiments. 

Constraints on milli-charged DM from direct detection experiments that we derived in this work are particularly pertinent due to recent interest to explain the observed EDGES signal by milli-charged DM~\cite{Munoz:2018pzp, Barkana:2018qrx, Berlin:2018sjs, Munoz:2018jwq}. Our result, in conjunction with the SENSEI limits~\cite{Crisler:2018gci}, are the first direct detection limits which constrain large values of charge $\epsilon$. 

We also derive the limits on hadronically-interacting DM-nucleon cross section with power-law velocity-dependence, i.e. $\sigma=\sigma_0\,v^n$. We consider the well-motivated range of velocity dependencies $n\in\{-4,-2,-1,0,1,2\}$~\cite{Holdom:1985ag, Chun:2010ve, Sigurdson:2004zp, ArkaniHamed:2008qn, Buckley:2009in,Tulin2013,Xingchen2018}. These results severely constrain hadronically-interacting DM by probing a broad range of underlying particle physics models. Particularly, for $n=-4$, we rule out the possibility that DM-hydrogen interaction can explain the EDGES signal for DM masses $\geq$ 1.25 GeV (see figure~\ref{XNDn4}), under the assumption that the relation between cross section on different nuclei is the one given by the Born approximation (eq.~\eqref{e_sigmavAF2}).

For the XQC experiment, we simultaneously weaken and strengthen the bounds in the literature. As discussed in appendix~\ref{ap:effth}, the efficiency with which nuclear recoil energy is thermalized was previously uncritically assumed to be 100\%, whereas it may be as small as $\epsilon_{\rm{th}}=0.02$ or less.  Until the thermalization efficiency factor has been measured experimentally for XQC, we adopt $\epsilon_{\rm{th}}=0.02$ as a potentially realistic estimate; this enlarges the ``hole''~\cite{nuDavis, Emken2018} in the parameter space of hadronically-interacting DM between CMB constraints~\cite{Dvorkin:2013cea,Gluscevic2017, Xu2018} and those from direct detections~\cite{nuDavis, bounds}; limits for $\epsilon_{\rm{th}}=0.1$ and 1 are also reported.  Limits for $\epsilon_{\rm{th}}=0.01$  are still weaker and will be reported elsewhere.  At the same time, we obtain more powerful limits from XQC by using a Monte-Carlo simulation to calculate the nuclear recoil spectrum valid also for multiple scatterings in the XQC detector. This enables us to extend the mass reach of XQC for $\epsilon_{\rm{th}}=0.02$ from 2 GeV to 420 MeV, for $\epsilon_{\rm{th}}=0.1$ from 1 GeV to 90 MeV, and for $\epsilon_{\rm{th}}=1$ from 225 MeV to 10 MeV.

The CSR experiment operating at the Earth's surface, was shielded by the Earth's atmosphere and the Earth's volume. We use a Monte-Carlo simulation to calculate the velocity distribution of the DM particles that can potentially trigger the CSR detector. We also consider the impact of thermalization efficiency in interpreting the CSR experiment. The SGED approximation~\cite{Starkman}, suitable in the limit of many interactions with small-deflection angles and essentially continuous energy-loss and critically examined in ref.~\cite{method}, is generalized for the case of power-law velocity-dependent contact interaction (see appendix~\ref{SGEDvel}) and milli-charged DM (see appendix~\ref{SGEDcol}).
\\

\noindent {\bf Acknowledgments:}  We thank Rennan Barkana, Jonathan Davis, Cora Dvorkin, Nadav J. Outmezguine, Linda Xu, and Xingchen Xu for providing the deatils of analyses in~\cite{Barkana:2018qrx,nuDavis,Xu2018,Xingchen2018}. We thank Paul Chaikin, Aditi Mitra, Paul Steinhardt and Andrew Wray for discussions about Frenkel pair production and thermalization processes in Silicon. MSM acknowledges support from the James Arthur Graduate Assistantship; the research of GRF has been supported by NSF-PHY-1212538 and AST-1517319.

\appendix

\section{Thermalization of low-energy nuclear recoils in Silicon} 
\label{ap:effth}

The XQC detector\cite{McCammon2002} was designed to measure the astrophysical diffuse X-ray background at high spectral resolution.  The detector consists of 34 Si micro-calorimeters operating at 60 mK, covered by a layer of HgTe to cause the X-rays to convert their energy to $e^\pm$ pairs which thermalize efficiently and produce a $\approx 5$ms spike in the temperature which is accurately measured by the thermometer.  XQC was calibrated with several X-ray sources in the keV range and the response is seen to be linear.  In the following, we examine whether it is a good assumption that the energy of a nucleus which recoils due to a DM collision, will fully thermalize as does the X-ray energy.  

The fractional energy loss per scattering of a DM on a nucleus at rest is 
\beq
f_{KE} = \left( \frac{2\,m\,m_A}{(m_A + m)^2 }\right) (1 - {\rm cos \, \xi_{CM}}). 
\eeq
For $m\ll m_A$, the mean fractional energy loss is $f_{KE} \sim 2\, m/m_A$, so the Si component of XQC is the most sensitive part of the target for $m_{DM} \lesssim$ few GeV. The maximum nuclear recoil energy given the escape velocity of the Galaxy is $\approx 2 \, (\frac{m}{2 m_p})^2$ keV and the mean value is $\approx 140 (\frac{m}{2 m_p})^2$ eV  In using XQC for constraining DM interactions, it has been assumed up to now (\cite{McCammon2002,Wandelt:2000ad} and subsequent works) that 100\% of the nuclear recoil energy is thermalized and measured by their quantum micro-calorimeter, when $E_{\rm nr}$ is above the 30 eV threshold.

For an illustrative recoil energy of 300 eV, the velocity of the a recoiling Si nucleus is a few 10's $\rm{km\cdot s^{-1}}$, while the typical speeds of electrons in Si atoms are 1000's $\rm{km\cdot s^{-1}}$.  Thus the nuclear recoil process is adiabatic as far as the atom is concerned, and the electron cloud is not disrupted -- the entire Si atom moves as a unit and electrons are not ionized\footnote{Higher energy nuclear recoils produce ionization.  Above $E_{\rm nr} \approx 3$ keV, the ionization signal is well-described by the Lindhard model \cite{Lindhard}, but measurements by DAMIC \cite{Chavarria:2016xsi} show that the ratio of ionization energy to the nuclear recoil energy in Si (the ionization ``quenching'' factor) is a strongly decreasing function of energy.  Moreover it is difficult to model:  already at $\approx 2$ keV the Lindhard model over-predicts the ionization signal by $\approx 50$\%, and this increases to a factor 2.5 at the lowest calibration point of DAMIC, $E_{\rm{nr}} \approx 680\,$eV, where the ionization signal is about 60 eV.}.  

The Si crystal is a covalently-bonded lattice, and when a Si atom is displaced, it migrates through the lattice leaving behind a vacancy and dislodging other Si atoms it encounters, losing roughly half its energy in each scattering\footnote{The actual differential cross section of atom-atom scattering in Silicon is slightly softer than hard-sphere, but this is sufficient for the present discussion.  It is modeled in \cite{Lindhard+AtomAtom63};' see the book \cite{vanLint+79} for a review and additional references.}.  Each collision produces another moving Si atom and associated vacancy, building up a cascade. The covalent bonds are weak and easily reorganized; the bonds of the crystal rearrange themselves in response to the moving atoms and eventually the system settles into a state with multiple vacancies and a corresponding number of displaced Si atoms called interstitials.  These are just Si atoms located at a position between the normal lattice sites; more precisely, these are called self-interstitials, to distinguish them from interstitial impurity atoms.  The Si lattice structure is fairly complicated, and there are numerous types of interstitial positions depending on where the atom is, relative to the lattice geometry. Such vacancy-interstitial pairs are called Frenkel pairs;  see, e.g., \cite{BarYamInterstitialPRL84,Leung_InterstitialPRL99,Rinke_InterstitialPRL09,Gusakov_FrenkelPair09}.   

The energy of a Frenkel pair in Si is $\approx5$ eV for the lowest-energy interstitial sites \cite{Gusakov_FrenkelPair09,Rinke_InterstitialPRL09}, but Frenkel pairs have a range of energies depending on the position of the interstitial in the lattice structure. The cascade ends when the kinetic energy of a given atom is too low to generate another Frenkel pair, and the remaining energy of each moving atom then thermalizes.  The cascade produced by a 300 eV nuclear recoil would produce $\approx 60$ Frenkel pairs and up to 6 ``generations''. The energy of motion of an interstitial is 0.1 eV \cite{Gusakov_FrenkelPair09,Rinke_InterstitialPRL09}, so $\approx 60$ interstitials could be expected to produce $\approx 6$ eV of thermal energy, for a 2\% thermalization efficiency. 
Note that in this estimation, the thermalization efficiency as a percentage of the initial recoil energy is just a simple percentage and would be applicable at other recoil energies.  A more detailed analysis of phonon production accompanying this process is left for future research.  In time, some interstitials and vacancies recombine or form more stable structures, e.g., di-vacancies, however this annealing process takes seconds to days or more, so is not relevant to measuring the energy deposit of individual DM recoils.

The general phenomenon of nuclear recoils producing interstitial defects is called the Wigner effect, after Eugene Wigner who first recognized the phenomenon in material irradiated by neutrons, in the moderator of a reactor.  A tremendous amount of energy can be tied up this way;  up to  2 kJ/g has been recorded.  The Wigner effect played a role in the Windscale nuclear reactor fire in 1957. 

The cumulative effect on the material of the Frenkel pairs is called displacement damage.  Modeling the process of non-ionizing energy loss (NIEL) is an active research field, which has been extensively studied in connection with MeV neutrons from reactors and in connection with the high radiation intensity environment of Si detectors at the LHC.  Modeling is also needed to calculate the effects of the radiation environment in space.   Experimental verification is mostly indirect, and so-far is mostly restricted to energies $\gtrsim 1$ MeV.  Direct measurement is problematic due to the difficulty of producing and working with mono-energetic low energy neutrons.   For a review of displacement damage see \cite{Srour+DisplacementDamageReview03};  for further information on NIEL and LHC applications, see  the CERN/Hamburg thesis \cite{Honniger07};  for a discussion of the implications for detectors in the high-luminosity LHC environment see \cite{Junkes11}.   For a microscopic model of the NIEL process see \cite{Huhtinen02};  the lowest energy case considered is a 1 MeV neutron and a 50 keV Si, but only model results are presented, without any data comparison.  These articles are by no means a complete listing of the literature, but suffice to demonstrate that a) modeling is very complex and involves much poorly known physics and materials science and b) modeling is only tested in regimes not applicable to the very low energy deposits relevant for the application of XQC  as a DM detector.

The XQC detector must be calibrated with a neutron source, to measure the thermalization efficiency of nuclear recoil energy.  The study by \cite{Chavarria:2016xsi} using radioactive neutron sources to measure the ionization yield in DAMIC illustrates the experimental approach.  Ref.  \cite{Chavarria:2016xsi} reports the quenching factor in Si at 12 different nuclear recoil energies, 0.68 -- 2.2 keV.  Even though the minimum energy is above the range of greatest interest for XQC, measuring the XQC thermal response for the DAMIC energies would greatly clarify whether lack of thermalization may be a problem for using the XQC technique for DM searches.  A more powerful approach could be to use the Spallation Neutron Source at ORNL.

Sapphire, the semi-conductor used in the CRESST detector, has many similarities with silicon including interstitial defects, so consideration should be given to how the CRESST device works, and measuring the efficiency factor for CSR as well. 

\section{Earth's crust model }\label{earth_crust_model}
We use the Preliminary reference Earth model (PREM)~\cite{PREM} to model the Earth's crust. We only need to consider 106.7 meters of the Earth's crust as DAMIC (at $z_{det}=106.7$ m underground), among underground experiments, gives the strongest limits on hadronically-interacting DM. The mass abundances of this layer of the Earth, with constant mass density $\rho_{crust}= 2.7\,\rm{gr\cdot cm^{-3}}$, are listed in table~\ref{t_earth_mass}.
\begin {table}[h!] 
\begin{center}	
	\begin{tabular}{||c|c|c||} 
		\hline
		Element& Atomic number& Mass fraction (\%)\\ [0.1ex] 
		\hline\hline
		O& 16& 46.6\\ 
		\hline
		Si& 28& 27.7\\ 
		\hline
		Al& 27&8.1 \\
		\hline
		Fe& 56&5.0 \\
		\hline
		Ca& 40&3.6 \\
		\hline
		Na& 23&2.8 \\
		\hline
		K& 39&2.6 \\
		\hline
		Mg& 24&1.5 \\
		\hline
		Total& --&97.9 \\
		\hline
	\end{tabular}
\end{center} 
\caption {The mass fraction of the most abundant elements in the Earth's crust for 0 -- 106.7 meters. }\label{t_earth_mass}
\end {table}
\section{Atmosphere model}\label{atm_model}
We use the U.S. Standard Atmosphere (USSA) 1976~\cite{USSA} to model the Earth's atmosphere. Up to 84.8 km, the atmosphere is divided into seven layers of ideal gas, each identified by a constant temperature gradient (lapse rate $L_m$). 
\par
Assuming the atmosphere to be in hydrostatic equilibrium and constant lapse rate $L_m$, atmosphere mass density in the m-th layer is
\beq
\label{eatmrho}
\rho_{atm}(z)=\rho_{atm}(z=z_m)\times\begin{cases}
	\left( 1+ \frac{L_m}{T_m}(z-z_m)\right)^{-(1+g/L_m\,R)} \hspace{1cm} L_m \neq 0\vspace{0.5cm}\\
	\exp\left( -\frac{(z-z_m)\,g}{R\,T_m}\right)  \hspace{3.3cm} L_m = 0\;,
\end{cases}
\eeq
where $T_m$ is the atmosphere's temperature at the bottom of the m-th layer, $z_m$. $R=$ 287.05 J/kg$\cdot$K is the gas constant for air.
\par
Knowing the sea level atmosphere mass density $\rho_{atm}(0)= 1.225\times10^{-3}\rm{gr\cdot cm^{-3}}$ and the sea level temperature $T_0=288.15$ K, the atmosphere's mass density can be calculated as a function of altitude. Table~\ref{t_atm_rho} listed this function for each layer of the Earth's atmosphere.
\begin {table}[t]
	\begin{center}	
		\begin{tabular}{||c|c|c|c|c|c|c||} 
			\hline
			Layer&$z_m$ (km)& $L_m$ (K/km) & $T_m$ (K)& $\frac{\rho(z=z_m)}{\rho(z=0)}$&$\frac{\rho(z)}{\rho(z=0)} $\\ [0.1ex] 
			\hline\hline
			0& 0& -6.5 & 288.15&1&$(1-z/44.31)^{4.256}$\\ 
			\hline
			1& 11& 0 &216.5& 0.297076&$0.297\exp((11-z)/6.34)$\\ 
			\hline
			2& 20&1.0&216.5&0.0718652&$(0.978+z/201.02)^{-35.16}$ \\
			\hline
			3& 32&2.8&228.5&0.0107959&$(0.857+z/57.94)^{-13.2}$\\
			\hline
			4& 47&0&270.5&1.165$\times10^{-3}$&$1.165\times10^{-3}\exp((47-z)/7.92)$\\ 
			\hline
			5& 51&-2.8&270.5&7.03$\times10^{-4}$&$(0.8-z/184.8)^{11.2}$\\ 
			\hline
			6& 71&-2&214.5&5.2$\times10^{-5}$&$(0.9-z/198.1)^{16.08}$\\
			\hline
			7& 84.8&---&186.8&5$\times10^{-6}$&---\\
			\hline
		\end{tabular}
	\end{center} 
	\caption{The mass density of the seven layers of the Earth's atmosphere as a function of altitude.}\label{t_atm_rho}
\end{table}
\par
The total over-head column depth at 84.8 km above the sea level is $\sim$0.00038\% of the total over-head column depth at the sea level (see table~\ref{t_atm_chi}). So, in this study we only consider this seven layers of the atmosphere and safely ignore upper layers. 
\begin {table}[t]
\begin{center}	
	\begin{tabular}{||c|c|c|c||} 
		\hline
		Layer&$z_m$ (cm)&  $\frac{\chi_h(z)}{\rho(z=0)} $ (cm)&$\chi_h(z)$ (g$\cdot\rm{cm^{-2}}$)\\ [0.1ex] 
		\hline\hline
		0& 0& $8.43\times10^5(1-z/4.43\times10^6)^{5.256}+524.89$&1033.36\\ 
		\hline
		1& 1.1$\times10^{6}$& $1.88\times10^5\exp(1.73-z/6.34\times10^5)+368.76$&231.11\\ 
		\hline
		2& 2$\times10^{6}$&$5.88\times10^5(0.978+z/2.01\times10^7)^{-34.16}-45.19$&56.23 \\
		\hline
		3& 3.2$\times10^{6}$&$4.75\times10^5(0.857+z/5.794\times10^5)^{-12.2}+12.36$&8.86\\
		\hline
		4&4.7$\times10^{6}$&$9.22\times10^{2}\exp(5.93-z/7.92\times10^5)+12.77$&1.14\\ 
		\hline
		5& 5.1$\times10^{6}$&$1.51\times10^6\,(0.8-z/1.848\times10^7)^{12.2}-1.166$&0.697\\ 
		\hline
		6& 7.1$\times10^{6}$&$1.16\times10^6\,(0.9-z/1.98\times10^7)^{17.08}$&0.040\\
		\hline
		7& 8.48$\times10^{6}$&---&0.0038\\
		\hline
	\end{tabular}
\end{center} 
\caption{The over-head column depth of the seven layers of the Earth's atmosphere as a function of altitude.}\label{t_atm_chi}
\end{table}
\par
The sea level mass abundances of the atmosphere's constituent are listed in table~\ref{t_atm_mass}.
\begin {table}[h!] 
\begin{center}	
	\begin{tabular}{||c|c|c||} 
		\hline
		Element& Atomic number& Mass fraction (\%)\\ [0.1ex] 
		\hline\hline
		N& 14& 78.08\\ 
		\hline
		O& 16& 20.94\\ 
		\hline
		Ar& 40&0.93 \\
		\hline
		Total& --&99.95 \\
		\hline
	\end{tabular}
\end{center} 
\caption {The sea level mass fraction of the most abundant elements in the Earth's atmosphere. }\label{t_atm_mass}
\end {table}

\section{Generalization of the DM{\scriptsize ATIS} code}\label{DMATIS_mod}
The DM{\scriptsize ATIS} code~\cite{code} was developed to simulate the propagation of DM particles with velocity-independent DM-nucleon cross section in the Earth's crust with constant mass density above the DAMIC experiment~\cite{bounds,method}. To model DM propagation in the Earth's  atmosphere with strong altitude dependence of its mass density, we generalized the code by 
\begin{itemize}
	\item Taking into account the geometry of overburden.
	\item Modeling the Earth reflection of hadronically-interacting DM particles.
	\item Changing the sampling variable from path length to column depth to determine the position of the next scattering. 
\end{itemize}
\subsection{Hybrid method to calculate the position of the next scattering in the atmosphere}\label{atm_analytic}
For a DM particle with mass $m$, velocity $v$ and cross section $\sigma_0$, the column depth can be sampled from the exponential distribution given in eq.~\eqref{echiS}. In this appendix, given the value of sampled column depth $\chi_s$, we present an analytic relation to find the altitude of a DM particle which was at altitude $h_{i-1}$ in the atmosphere.
\begin{figure}[tbp]\centering
	\includegraphics[width=0.65\textwidth]{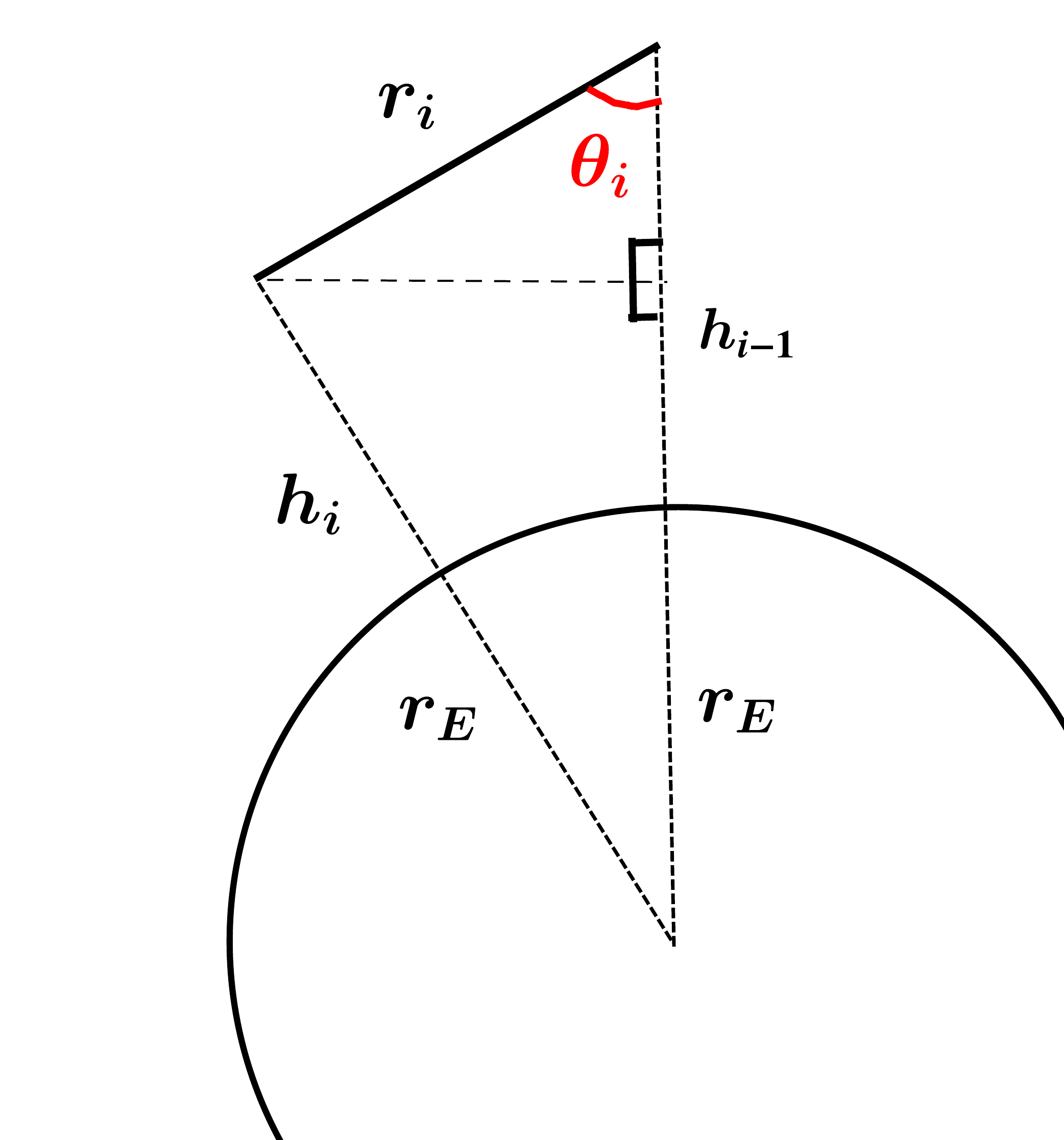}
	\caption{\label{scattering_geo} Schematic drawing of the DM scattering in the Earth's rest frame.}
\end{figure} 
The path length in the $i$-th step of a DM particle start traveling along zenith angle $\theta_i$ from altitude $h_{i-1}$ to reach altitude $h_i$ is (see figure~\ref{scattering_geo})
\beq
\label{r_i}
r_i=\cos\,\theta_i\,(R_E+h_{i-1})\pm\sqrt{(R_E+h_{i-1})^2(\cos^2\theta_i-1)+(R_E+h_{i})^2}\;,
\eeq
where $R_E=6.37\times10^3$ km is the Earth's radius. The path length infinitesimal changes can be related to infinitesimal changes in the final altitude by
\beq
\label{drdh}
\frac{dr_i}{dh_i}=\pm\left( (\cos^2\theta_i-1)\left( \frac{R_E+h_{i-1}}{R_E+h_{i}}\right) ^2+1\right)^{-1/2} \;.
\eeq
Using eq~\eqref{drdh}, the accumulated column depth along the path length $r_i$ is
\beq\begin{split}
	\label{chi_r}
	\chi(r_i)&=\int_0^{r_i}\rho(z)\,dr\\
	&=\int_{h_i}^{h_{i-1}}\rho(z)\left( (\cos^2\theta_i-1)\left( \frac{R_E+h_{i-1}}{R_E+z}\right) ^2+1\right)^{-1/2}  dz\;.
\end{split}
\eeq
For altitudes of the atmosphere's first seven layers ($<$ 84.8 km), $ R_E+z\simeq R_E+h_{i-1}$ and the geometry of the Earth can be neglected in the calculation of the accumulated path length
\beq\begin{split}
	\label{chi_r_appr}
	\chi(r_i)&\simeq\frac{1}{\cos\,\theta_i}\left\lbrace \int_{h_i}^{\infty}\rho(z)dz-\int_{h_{i-1}}^{\infty}\rho(z)dz\right\rbrace \\
	&=\frac{1}{\cos\,\theta_i}\left\lbrace \,\chi_h(h_i)-\chi_h(h_{i-1})\,\right\rbrace\;,
\end{split}
\eeq
where $\chi_h(h_i)$ is the over-head column depth at altitude $h_i$. \\
Using $\chi(r_i)\equiv\chi_s$ in eq.~\eqref{chi_r_appr}, altitude $h_i$ can be calculated analytically by
\beq\begin{split}
	\label{chi_h_inv}
	h_i=\chi_h^{-1}(\cos\,\theta\,\chi_s+\chi_h(h_{i-1}))\;,
\end{split}
\eeq
where $\chi_h^{-1}$ is the inverse function of the over-head column depth functions given in table~\ref{t_atm_chi}. 
\par
\begin{figure}[tbp]\centering
	\includegraphics[width=0.7\textwidth]{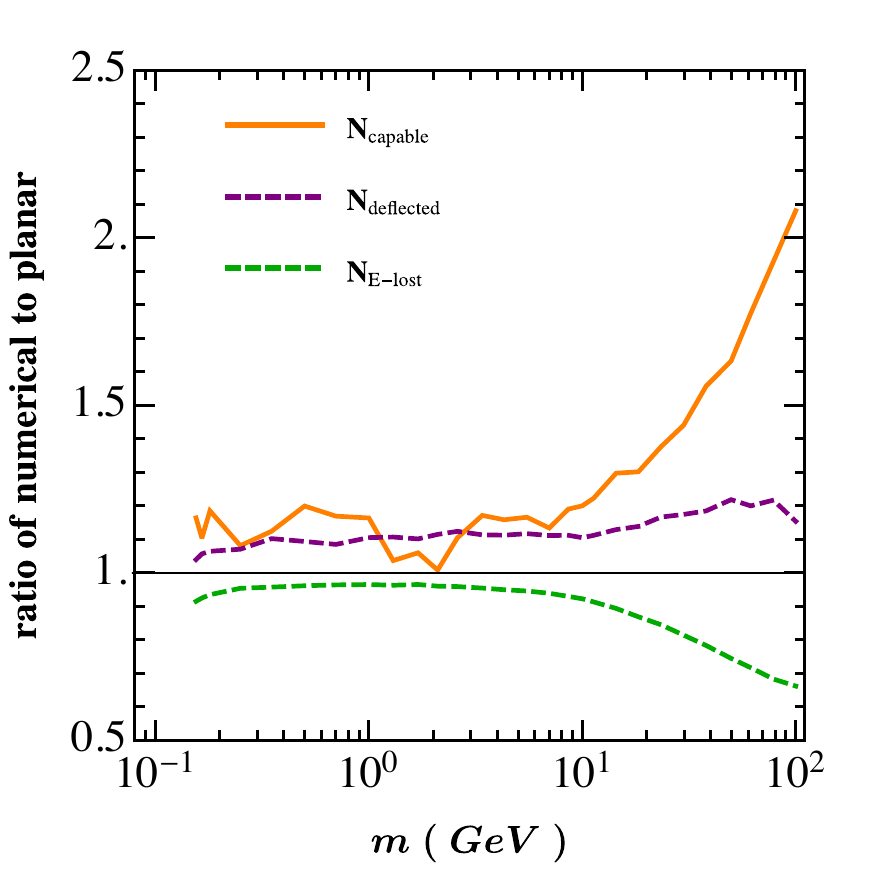}
	\caption{\label{Hybrid_to_Planar} The ratio of the capable number of events (orange solid line) calculated at the 90\% CL upper reach of the CRESST 2017 surface run (CSR) experiment on the velocity-independent DM-nucleon cross section using the hybrid method (used in this work) to the result of the Planar approximation. The planar approximation also underestimates/overestimates the number of deflected DM particles/the number of DM particles that their energy falls below the threshold of the  before reaching the detector.}
\end{figure}

Eq.~\eqref{chi_h_inv} enables us to calculate altitude $h_i$ accurately for $|\cos\theta_i|>0.16$. For $|\cos\theta_i|\leq0.16$, we numerically integrate the atmosphere's density along the path length using eq.~\eqref{chi_r}. Figure~\ref{Hybrid_to_Planar} shows the ratio of the capable number of events calculated by hybrid method to the result of the Planar approximation. This shows that using the Planar approximation underestimates the number of capable DM particles by a factor as large as 2. This can be understood by noticing that the number of deflected particles is underestimated in the planar approximation due to overestimation of the column-depth for particles scattering along zenith angles $\theta_i\simeq\pi/2$. 

\subsection{Earth, a mirror of hadronically-interacting DM particles}\label{earth_reflection_effect}
\begin{figure}[tbp]\centering
	\includegraphics[width=0.99\textwidth]{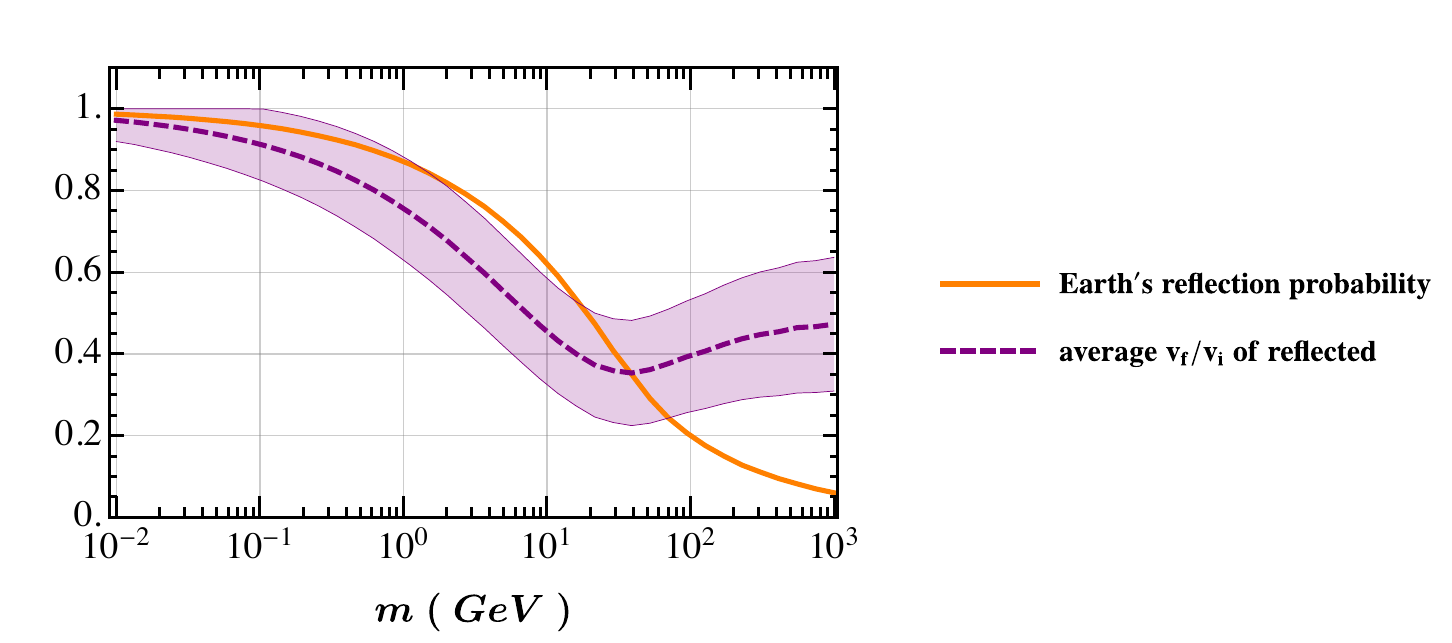}
	\caption{\label{Earth_reflection} The Earth reflection probability (solid orange) and the average ratio of final to initial velocities of the reflected DM particles (dashed purple) as a function of DM mass. The purple band shows one standard deviation of the ratio of final to initial velocities of the reflected DM particles. This result is independent of the value of DM-nucleon cross section as long as the cross section is big enough to cause DM particles to scatter multiple times in the Earth. }
\end{figure}

Hadronically-interacting particles that enters the Earth can scatters back to the Earth's atmosphere~\cite{Kavanagh:2016pyr}. Figure~\ref{Earth_reflection} shows the Earth reflection probability and average ratio of final to initial velocities of the reflected DM particles as a function of DM mass. This result is governed by simulation of DM particles entering the Earth's crust and tracking their position and velocity until 
\begin{itemize}
	\item They scatter back to the Earth's atmosphere, or
	\item Their velocity fall below the Earth's escape velocity 11.2 $\rm{km\cdot s^{-1}}$.
\end{itemize}
\par
Figure~\ref{Earth_reflection} shows that light DM particles, which have a larger probability of reflection before capture, lose a small fraction of their velocity in the Earth. So, a DM particle can cause more than one event in the CSR detector which operated on the Earth's ground. The Earth reflection of DM particles increases the rate of expected number of events for CSR and reduces the value of the CSR maximum DM-nucleon cross section reach.
\par
The DM{\scriptsize ATIS} code for calculating the expected number of events of CSR is modified to model this effect by tracking DM particles as they may scatter in and out of the Earth after propagating through the Earth's atmosphere, before and/or after interacting in the detector, until their energy falls below the minimum energy to be able to deposit 19.7 eV. 

\subsection{Column depth instead of path length}\label{vel_generalization_of_DMATIS}
The probability that a DM particle of mass $m$ and velocity $v$ does not scatter off target nuclei of mass number $A$ in $[0,r+dr]$ is
\beq
\label{evdQA}
Q_A(r+dr,v)=Q_A(r,v)\,\left( 1-\frac{dr}{\lambda_A(r, v)}\right) \,,
\eeq
where $\lambda_A(r, v)\equiv(n_A(r)\,\sigma_A(v))^{-1}$ is the mean path length of DM particles with velocity $v$ at position $r$. Correspondingly, the probability that a DM particle does not scatter off any nucleus in a target with a mix of nuclei is
\beq
\label{evdQ}\begin{split}
	Q(r+dr,v)&=Q(r,v)\,\left( 1-\sum_A\frac{dr}{\lambda_A(r, v)}\right) \\
	&=Q(r,v)\,\left( 1-\frac{dr}{\lambda_{eff}(r, v)}\right)\,,
\end{split}\eeq
where $\lambda_{eff}(r, v)\equiv(\sum_A\lambda_A^{-1}(r,v))^{-1}$ is the mean path length of DM particles in target with a mix of nuclei. Solution of eq.~\eqref{evdQ} is 
\beq
\label{evQ}\begin{split}
	Q(r,v)&=\exp\left( -\sum_A\sigma_A(v)\int_{0}^{r}n_A(r')dr'\right) \\
	&=\exp\left(- \sum_A\frac{\sigma_A(v)\,\chi_A(r)}{m_A}\right) \,,
\end{split}\eeq
where $\chi_A(r)=\int_{0}^{r}\rho_A(r')dr'$ is the column depth of target nuclei of mass number $A$ accumulated from $r'=0$ to $r'=r$. For a target with position independent mass composition, i.e. $f_A(r')=f_A$, eq.~\eqref{evQ} is simplified
\beq
\label{evQ2}\begin{split}
	Q(r,v)&=
	\exp\left(- \chi(r)\sum_A\frac{\sigma_A(v)\,f_A}{m_A}\right)\\
	&=\exp\left( -\frac{\chi(r)}{\chi_{eff}(v)}\right) \,,
\end{split}\eeq
where $\chi(r)=\int_{0}^{r}\rho_T(r')dr'$ is the column depth accumulated over path length $r$ and $\chi_{eff}(v)\equiv\left( \sum_A\frac{\sigma_A(v)\,f_A}{m_A}\right)^{-1} $ is the mean column depth in a target with a mix of nuclei.
\par
With a change of variables from $(r,v)$ to $(\chi,\chi_{eff})$, the probability that a DM scatters off any nucleus in a target with a mix of nuclei in [$\chi, \chi+d\chi$] is 
\beq
\label{evPdchi}\begin{split}
	P(\chi,\chi_{eff})\,d\chi&=\frac{d\chi}{\chi_{eff}(v)}\,Q(\chi,\chi_{eff})\, ,
\end{split}\eeq
and therefore
\beq
\label{evP}\begin{split}
	P(\chi,\chi_{eff})&=\frac{1}{\chi_{eff}(v)}\exp\left( -\frac{\chi(r)}{\chi_{eff}(v)}\right)\,.
\end{split}\eeq
\par
In a Monte-Carlo simulation of DM particles with velocity-dependent cross section $\sigma_A(v)$ traveling through a shielding material with position dependent mass density $\rho(r)$, the column depth should be sampled from the probability distribution given in eq.~\eqref{evP}, instead of path length in the case of velocity-independent and target position-independent mass density
\beq
\label{echiS}\begin{split}
	\chi_s = -\chi_{eff}(v) \ln(1-ran_i)\,,
\end{split}\eeq
where $ran_i\in[0,1]$ is a random number representing the column depth's cumulative probability. To determine the position of the next scattering $r_s$, the column density is numerically accumulated along the current DM velocity vector until it reaches the sampled column depth $\chi_s$. The mass number of the target nuclei in the next scattering will be determined based on the mass composition of the shielding material at position $r_s$.

\section{Generalization of the SGED method}\label{SGEDgen}
The SGED method, which was proposed by~\cite{Starkman}, calculates the maximum velocity-independent DM-nucleon cross section for which a given detector would be able to see any events at all due to significant energy-loss of hadronically-interacting DM particles through their scatterings in an overburden. \cite{method} improved this crude SGED approximation by using the number of events observed by the experiment rather assuming none are observed and showed that this modification improves the SGED cross section limits for DAMIC by 15\%. In this appendix, we extend the crude SGED method to find the maximum cross section for the case where a) mediator is much heavier than the DM momentum transfer and has a DM-nucleus power-law velocity-dependency b) milli-charged DM.
\subsection{Heavy mediator in the Born approximation}\label{SGEDvel}
In the commonly-used Born approximation, in the case of a heavy mediator for which mediator mass is much greater than the typical DM momentum transfer, $\mu_A\,v\approx10^{-3}\mu_A$, the differential DM-nucleus cross section in scattering off a nucleus of mass number $A$ is
\beq\begin{split}
	\label{edsAdEr_con}
	\frac{d\,\sigma_A}{d\,E_{\rm{nr}}}&=\,\frac{m_A\,\sigma_A(v)}{2\,\mu^2_A\,v^2}  \;,
\end{split}
\eeq
where $m_A$ is nucleus mass and $\mu_A$ is DM-nucleus reduced mass. Here we ignore the nuclear form factor as we are interested in DM masses $\lesssim$100 GeV. 

Due to recoil energy independence of the expression in the right-hand-side of eq.~\eqref{edsAdEr_con}, DM scatterings of nuclei in this regime are isotropic. And, the spin-independent DM-nucleus cross section is related to DM-nucleon cross section, $\sigma_p(v)$, by
\beq\begin{split}
	\label{esigmaAE}
	\sigma_{A}(v)&=\sigma_p(v) \left(\frac{\mu_A}{\mu_p} \right) ^2A^2   \;,
\end{split}\eeq
where $\mu_p$ are the DM-nucleon reduced mass. 
\par 
The differential energy-loss of a DM particle with mass $m$ and energy $E$ to nuclei of mass number $A$ while passing through a shielding material is
\beq
\begin{split}\label{edEAE}
	\frac{dE_A}{dz}&=-\frac{\left\langle E_{\rm{nr},A}\right\rangle}{\lambda_{eff}(z, v)}\;,
\end{split}\eeq
where $\lambda_{eff}(z, v)\equiv(\sum_{A}n_A(z)\sigma_A(v))^{-1}$ is the mean path length at distance $z$ of DM particles with energy $E$ in a shielding material with a mix of nuclei. $n_A(z)=\rho_A(z)/m_A$ is the number density of nuclei of mass number $A$ with the mass density $\rho_A(z)$ at distance $z$. 
\par
$\left\langle E_{\rm{nr},A}\right\rangle$, the average nuclear recoil energy of the DM particle scattering off a nucleus of mass number $A$, is
\beq
\begin{split}\label{eErAve}
	\left\langle E_{\rm{nr},A}\right\rangle = \frac{2\,\mu_A^2}{m\,m_A}\,E\;.
\end{split}\eeq
\par
Using eqs.~\eqref{eErAve}, ~\eqref{esigmaAE}, and~\eqref{eErAve}, the total differential energy-loss of a DM particle in a scattering off a shielding material with a mix of nuclei is 
\beq
\begin{split}\label{edEdzE}
	\frac{dE}{dz}&=-E\sum_{A}n_A(z)\,\sigma_A(v)\left( \frac{2\,\mu_A^2}{m\,m_A}\right) \\&=-\sigma_p(v)\,\frac{2\,E}{m}\sum_A \rho_A(z)\left( \frac{\mu_A^2}{m_p\,\mu_p}\right)^2\\&=-\,\sigma_0\,v^{n+2} \sum_A \rho_A(z)\left( \frac{\mu_A^2}{m_p\,\mu_p}\right)^2\;,
\end{split}\eeq
where in the last expression, the power-law velocity-dependent DM-nucleus cross section of the form $\sigma_p(v)\equiv\sigma_0\,v^n$ is substituted.
\par
For $n\neq0$, the velocity of a DM particle after traveling a distance $L$ in the continuous energy-loss approximation is
\beq
\label{evEL}
v(L)^{-n}= v(0)^{-n}\, + \frac{2\,n\,\sigma_0}{m}\sum_A \chi_A(L)\left( \frac{\mu_A^2}{m_p\,\mu_p}\right)^2 \;,
\eeq
where $v(0)$ is the initial velocity of the DM particle before entering the shielding material and $\chi_A(L)$ is the overhead column depth of nuclei of mass number $A$ from $z=0$ to $z=L$:
\beq
\label{eChi180}
\chi_A(L)=\int_{0}^{L}\rho_A(z)\,dz\;.
\eeq
\par
Following the simplification of the crude SGED method in assuming that all particles above the overburden have the maximum velocity of $v(0)=v_{\rm{max}}$, the maximum cross section for which the detector can potentially be sensitive is
\beq
\label{evSGEDb}
\sigma_0^{\rm{max}}=\frac{m}{2\,\sum_A \,\chi_A(L) \left( \frac{\mu_A^2}{m_p\,\mu_p}\right)^2} \left(\frac{v_{\rm{min}}^{-n}(A')-v_{\rm{max}}^{-n}}{n}\right) \;,
\eeq
where $v_{\rm{min}}(A')$ is the minimum velocity of a DM particle to scatter a target nucleus of mass number $A'$ in the detector\footnote{ We distinguish by a prime notation ($'$) between nuclei in the detector and nuclei in the overburden.} and produce the threshold recoil energy $E_{\rm{nr}}^{th}$
\beq
\label{e_v'm}
v_{\rm{min}}(E_{\rm{nr}},A')=\sqrt{\frac{m_{A'}\,E_{\rm{nr}}^{th}}{2\,\mu^2_{A'}}}\;.
\eeq
\par
Instead of using eq.~\eqref{evSGEDb}, we use eq.~\eqref{evEL} to calculate the velocity distribution of DM particles at the detector, $f(\vec{v},\vec{v}_{det},\sigma_0)$. Using the velocity distribution, the expected differential number of events for power-law velocity dependent cross section of the form $\sigma_p=\sigma_0\,v^{n}$ in a detector of nuclei of mass number $A'$ is
\beq\begin{split}
	\label{evdNdEr}
	\frac{dN_{A'}}{dE_{\rm{nr}}}(\sigma_0)=&\,t_e\,M_{A'}\,A'^2\,\frac{\rho\,\sigma_0}{2\,m\,\mu^2_p}F_{A'}^2(E_{\rm{nr}})\int_{v_{\rm{min}}(E_{\rm{nr}},A')} v^{n-1}\,f(\vec{v},\vec{v}_{det},\sigma_0)\,d^3v\;.
\end{split}\eeq
\par
As the trial cross section is decreased from some large value, the expected differential number of events, calculated using eq.~\eqref{evdNdEr}, is increasing monotonically. The improved SGED 90\% lower bound on the allowed cross section would then be defined to be the value of $\sigma_0$ for which the expected total number of events equals the 90\% CL upper limit of the observed total number of events.
\subsection{Milli-charged DM}\label{SGEDcol}
As discussed in the text, milli-charged DM particles favor forward-scatterings. This makes calculation of the velocity distribution of the milli-charged hadronically-interacting DM particles at the detector computationally cumbersome. In contrast to a hadronically-interacting DM with isotropic scattering in the CM frame, a milli-charged hadronically-interacting DM (independent of its mass):
\begin{itemize}
	\item Favors forward-scattering in the lab frame due to having smaller scattering angles in the CM frame.
	\item Has larger number of scatterings in the overburden due to smaller average energy losses in the each scattering.
\end{itemize}
These two make a SGED type approximation more accurate as discussed in~\cite{bounds}. Here, we derive an analytic expression for the energy-loss of a milli-charged hadronically-interacting DM in an overburden.

The differential DM-nucleus cross section of a milli-charged DM with velocity $v$ is:
\beq\begin{split}
	\label{edsAdEr}
	\frac{d\,\sigma_A}{d\,E_{\rm{nr}}}&=\,\frac{2\,\pi\,Z_A^2\,\epsilon^2\alpha^2}{m_A\,v^2\,E_{\rm{nr}}^2}  \;,
\end{split}
\eeq
where $\epsilon\,e\equiv\epsilon_{\rm{th}}\,g_D$ is the electric charge of a DM particle which is coupled to dark photon with charge $g_D$. $m_A$ is the mass of nucleus, and $Z_A$ is the charge number of the nucleus. Here we ignore the nuclear form factor as we are interested in DM masses $\lesssim$100 GeV.

As it is pointed out in~\cite{Foot:2003iv}, the differential energy-loss of a milli-charged hadronically-interacting DM particle with mass $m$ and energy $E$ to nuclei of mass number $A$ while passing through a shielding material is
\beq
\begin{split}\label{edEAdz}
	\frac{dE_A}{dz}&=-\frac{\rho_A(z)}{m_A}\int_{E_{\rm{nr},A}^{\,\rm{max}}}^{E_{\rm{nr},A}^{\,\rm {screen}}}E_{\rm{nr},A}\,\frac{d\,\sigma_A}{d\,E_{\rm{nr},A}}\,dE_{\rm{nr},A}\\&=-2\,\pi\rho_A(z)\,\left( \frac{Z_A\,\epsilon\,\alpha}{m_A\,v}\right)^2\,\ln\left( \frac{E_{\rm{nr},A}^{\rm{\,max}}}{E_{\rm{nr},A}^{\,\rm {screen}}}\right)  \;,
\end{split}\eeq
where $E_{\rm{nr},A}^{\,\rm{max}}=(2\,\mu_A\,v)^2/2\,m_A$ is the maximum recoil energy. $E_{\rm{nr},A}^{\,\rm {screen}}\approx(\alpha\,m_e)^2/2\,m_A$ is the minimum recoil energy due to screening of the nucleus over distance $\sim(\alpha\,m_e)^{-1}$.

By substituting $E=\frac{1}{2}m\,v^2$ and rearranging eq.~\eqref{edEAdz}, the following differential equation is found
\beq
\begin{split}\label{edvdz}
	\frac{v^2\,d\,v^2}{\ln\,(\eta_A^2\,v^2)}=-\,\pi\left(\frac{2\,Z_A\,\epsilon\,\alpha}{m_A} \right)^2\,\frac{\rho_A(z)}{m}\,d\,z \;,
\end{split}\eeq
where $\eta_A=2\,\mu_A/\alpha\,m_e$.
 
 Noticing $\int v^2\,dv^2\,/\ln(\eta_A^2\,v^2)=Ei(4\ln(\eta_A\,v))/\eta_A^4$, eq.~\eqref{edvdz} can be integrated analytically to calculate final velocity $v_f$ of a DM particle after passing through an overburden to reach $z=L$
 
 \beq
 \begin{split}\label{evcol}
 	Ei\left( 4\,\ln\left( \eta_A\,v_f\right) \right) = Ei\left( 4\,\ln\left( \eta_A\,v\right) \right)-\,\pi\left(\frac{2\,\eta_A^2\,Z_A\,\epsilon\,\alpha}{m_A} \right)^2\,\frac{\chi_A(L)}{m}\;,
 \end{split}\eeq
where $Ei(x)\equiv\int_{-x}^{\infty}e^{-t}/t\, dt$ is the exponential integral function. Expansion of the inverse of the exponential integral function in terms of Chebyshev polynomials~\cite{1986BAICz..37....8P} is used to calculate the velocity of hadronically-interacting DM particles $v_f$ after passing an overburden. 

Using the unit-normalized velocity distribution of hadronically-interacting DM at the detector $f(\vec{v},\vec{v}_{det},\epsilon)$, the expected differential number of events in a detector made of nuclei of mass number $A'$ \footnote{ We distinguish by a prime notation ($'$) between nuclei in the detector and nuclei in the overburden.} is

\beq\begin{split}
	\label{edNdErcol}
	\frac{dN_{A'}}{dE_{\rm{nr}}}(\epsilon)=&\,t_e\,N_{A'}\,\frac{\rho}{m}\int_{v_{\rm{min}}(E_{\rm{nr}},A')} \frac{d\,\sigma_{A'}}{d\,E_{\rm{nr}}}\,v\,f(\vec{v},\vec{v}_{det},\epsilon)\,d^3v\;,
\end{split}\eeq
where $N_{A'}\equiv M_{A'}/m_{A'}$ is the number of nuclei of mass number $A'$ in the target. 
\section{XQC detector spectrum}\label{XQC_detail}
The XQC~\cite{McCammon2002} calorimeters were composed of a 0.96 $\mu$m film of HgTe mounted on a 14 $\mu$m substrate of Si. The total mass of each component of the XQC detector is
\beq
\label{e12}
M_A=\begin{cases}
	1.30\times10^{-3}\,g\hspace{1cm} A=Si\\
	1.02\times10^{-4}\,g  \hspace{1cm} A=Te\\
	1.61\times10^{-4}\,g \hspace{1cm} A=Hg\;.
\end{cases}
\eeq
The exposure time of the experiment depended on the thermal response, $E_{\rm{T}}$, so we write $t_e(E_{\rm{T}})=100.7\,\mathrm{f_e}(E_{\rm{T}})$ seconds, where $\mathrm{f_e}(E_{\rm{T}})$ is the fraction of the exposure time that XQC was sensitive to thermal energy $E_{\rm{T}}$. Table \ref{t_XQC_spec_kappa} contains the sensitivity factors and the observed number of events in each bin taken from Erickcek et al.~\cite{Erickcek2007}. 
\par
\begin {table}[h!] 
\begin{center}	
	\begin{tabular}{||c|c|c|c||} 
		\hline
		i& $E^i_{\rm{T}}=\epsilon_{\rm{th}}^{-1}E_{\rm{nr}}^i\;(eV)$& $\mathrm{f_e}(E_{\rm{T}})$ & $O_i $\\ [0.1ex] 
		\hline\hline
		1& 29 -- 36& 0.38 & 0\\ 
		\hline
		2& 36 -- 128& 0.5 & 11\\ 
		\hline
		3& 128 -- 300&1&129 \\
		\hline
		4& 300 -- 540&1&80\\
		\hline
		5& 540 -- 700&1&90\\ 
		\hline
		6& 700 -- 800&1&32\\ 
		\hline
		7& 800 -- 945&1&48\\
		\hline
		8& 945 -- 1100&1&31\\
		\hline
		9& 1100 -- 1310&1&30\\
		\hline
		10& 1310 -- 1500&1&29\\
		\hline
		11& 1500 -- 1810&1&32\\
		\hline
		12& 1810 -- 2505&1&15\\
		\hline
		13& $\geq$4000&1&60\\
		\hline
	\end{tabular}
\end{center} 
\caption {The XQC sensitivity factor and the observed number of events in each thermal energy bin. Bin 2505 -- 4000 eV is ignored due to its contamination by the detector's interior calibration sources. }\label{t_XQC_spec_kappa}
\end {table}

We distinguish between the measured thermal energy $E_{\rm{T}}$ by XQC and the induced nuclear recoil energy induced by DM scattering $E_{\rm{nr}}$. In this work, we linearly scale the expected-to-be-measured nuclear recoil energy with a thermalization efficiency factor $\epsilon_{\rm{th}}$ to calculate the measured thermal energy; i.e. $E_T=\epsilon_{\rm{th}}\,E_{\rm{nr}}$.
\bibliography{vb}
\end{document}